\documentclass[10pt,a4paper]{article}
\usepackage{float}
\usepackage{graphicx} 
\usepackage[export]{adjustbox}
\usepackage[margin={1.5cm,2cm}]{geometry}
\usepackage{relsize}
\usepackage{doi}
\usepackage[numbers]{natbib}
\hypersetup{nolinks=true}
\usepackage{xcolor}
\usepackage{mathtools, nccmath}
\usepackage[font=small]{caption}
\usepackage{authblk}

\graphicspath{ {./images/} }

\newcommand{\myBigFig}[2]{
\begin{figure*}
\centering
\includegraphics[max width=0.8\textwidth]{#1}
\caption{#2}
\label{fig:#1}
\end{figure*}
}

\newcommand{\code}[1]{\smaller{}\texttt{#1}\larger{}}
\newcommand{\secRef}[1]{Section \ref{sec:#1}}

\newcommand{\tabRef}[1]{Table \ref{tab:#1}}
\newcommand{\figRef}[1]{Fig.\@ \ref{fig:#1}}

\newcommand{\pseudocodeKeyword}[0]{Pseudocode}
\newcommand{\codeRef}[1]{\pseudocodeKeyword{} \ref{code:#1}}
\newcommand{\supPseudocodeKeyword}[0]{Supplementary Pseudocode}

\floatstyle{ruled}
\newfloat{codeFloat}{tbp}{cod}
\floatname{codeFloat}{\pseudocodeKeyword{}}
\newfloat{supCodeFloat}{tbp}{cod}
\floatname{supCodeFloat}{\supPseudocodeKeyword{}}

\newcommand{\abs}[1]{\ensuremath\left\lvert#1\right\rvert}

\newcommand{\algorithmName}[0]{DOSA-MO}
\newcommand{\AlgorithmName}[0]{DOSA-MO}
\newcommand{\AlgorithmExtName}[0]{Dual-stage Optimizer for Systematic overestimation Adjustment in Multi-Objective problems}
\newcommand{\AlgorithmCodeName}[0]{DosaMO}
\newcommand{\paperTitle}[0]{Dual-stage optimizer for systematic overestimation adjustment applied to multi-objective genetic algorithms for biomarker selection}

\title{\paperTitle{}}
\author[1]{Luca Cattelani}
\author[1]{Vittorio Fortino\thanks{Corresponding Author: vittorio.fortino@uef.fi}}
\affil[1]{Institute of Biomedicine, School of Medicine, University of Eastern Finland, Kuopio, 70211, Finland.}
\date{}

\begin{document}

\bibliographystyle{unsrtnat}

\maketitle

\begin{abstract}
The challenge in biomarker discovery using machine learning from omics data lies in the abundance
of molecular features but scarcity of samples. Most feature selection methods in machine learning require evaluating 
various sets of features (models) to determine the most effective combination. 
This process, typically conducted using a validation dataset, involves testing different feature sets to optimize the model's performance.
Evaluations have performance estimation error and when the selection involves many models
the best ones are almost certainly overestimated. Biomarker identification with feature selection methods can be addressed as a multi-objective problem
with trade-offs between predictive ability and parsimony in the number of features. Genetic
algorithms are a popular tool for multi-objective optimization but they evolve numerous solutions thus are prone to
overestimation. Methods have been proposed to reduce the overestimation after a model has already been selected
in single-objective problems, but no algorithm existed capable of reducing
the overestimation during the optimization, improving model selection, or applied in the
more general multi-objective domain. We propose \algorithmName{}, a novel multi-objective optimization
wrapper algorithm that learns how the original estimation, its variance, and the feature set size of the solutions
predict the overestimation. \algorithmName{} adjusts the expectation of the performance during the optimization, improving the
composition of the solution set. We verify that \algorithmName{} improves the performance of a state-of-the-art genetic
algorithm on left-out or external sample sets, when predicting cancer subtypes and/or patient overall survival, using
three transcriptomics datasets for kidney and breast cancer.
\\~\\
\textbf{Data and source code availability.}
The original gene expression data used in this study is from public repositories.
The preprocessed data, source code, and detailed numerical results are available in a public server
(\url{github.com/UEFBiomedicalInformaticsLab/BIODAI/tree/main/DOSA_MO}).
\end{abstract}

\twocolumn

\section{Introduction}

Molecular biomarker discovery with machine learning (ML) is usually limited by data that includes many features but few samples \cite{li2022data}.
This renders trained models prone to overfitting and the evaluation prone to estimation error.
Hyperparameter tuning, including feature selection, is often a crucial aspect of the optimization process
where the model's performance is assessed using a training, validation, and test paradigm.
This essential aspect involves choosing the most effective subset of features, essentially optimizing the ML's model.
Selecting the best model from many often leads to overestimation, given by a significant gap between validation set performance and actual, real-world performance,
a phenomenon known as the ``winner's curse''.
The models that fit the noise present in the validation set are advantaged, a phenomena sometimes referred as overfitting on the validation set \cite{Cawley2010}.
This is also exacerbated by data scarcity. Seeking higher accuracies by expanding hyperparameter configurations can enhance model performance on validation sets.
However, this often results in heightened overestimation, leading to reduced or potentially negative impact on test set performance.

In biomarker discovery, the focus is often on optimizing the accuracy of machine learning models using the selected molecular features, while also minimizing the number of features to ensure clinical feasibility and resource efficiency. Characterising all the best compromises (or trade-offs) between predictive value and feature set size is a multi-objective (MO) optimization problem \cite{Deb2014MO,morales2023}, and it can be solved by means of MO feature selection (MOFS) techniques \cite{Serra2019,Fortino2020,Cattelani2022,Cattelani2023}. These techniques aim to identify not just a single best solution, as in single-objective (SO) problems, but rather a Pareto front of solutions. This front is the set of optimal solutions that illustrate the trade-offs between different objectives. However, all candidate solutions (or biomarker models selected by the employed MOFS technique) are evaluated on the validation set, which can result in the overestimation of the performance of the selected models.

K-fold cross-validation (CV) stands as the predominant methodology for machine learning assessment, with its advantages and limitations extensively explored, particularly in SO scenarios \cite{krstajic2014,Wong2015,Wong2020}.
K-fold CV returns a model trained on all the available samples and an estimation of its performance computed by averaging $k$ CV results. The obtained model performance is usually underestimated when only one hyperparameter configuration is used. However, it tends to be overestimated when multiple configurations are evaluated, and the model (or feature set) with the best performance is returned \cite{Cawley2010}. This situation becomes particularly pronounced in MO approaches, where identifying multiple favorable trade-offs often leads to an increased number of models to be evaluated.
A practical, illustrative example is presented in Cattelani et al.\@ Fig.\@ 1, where it is evaluated the performance of various gene expression-based molecular feature sets for classifying breast cancer subtypes \cite{Cattelani2022}.
In this MO setting, balanced accuracy and feature set size serve as evaluation metrics. The expected balanced accuracy tends to increase with the addition of more features.
Similarly, overestimation – quantified as the difference between training and test set performance (the gap between ``inner cv'' and ``test'' results in the figures) – also grows as the estimation improves.
These correlations may suggest that characteristics of the solutions might be predictive of the degree of overestimation.

Although various methods have been developed to enhance performance estimation in model selection using k-fold CV, their design and implementation have been limited to SO problems. Tsamardinos et al. \cite{tsamardinos2015} compared double CV, the Tibshirani and Tibshirani method \cite{tibshirani2009}, and nested CV in their ability to improve the estimation of the fitness for SO problems. These algorithms modify fitness estimations but do not change the model selection process; the chosen model remains the same as it would be using simple hyperparameter optimization with k-fold CV for model evaluation. Automated ML (AutoML) tools offer an approach designed to explore various model and hyperparameter combinations. These tools aim to identify and deliver the most effective model along with an assessment of its performance. In Tsamardinos et al.\@ six AutoML tools were compared \cite{tsamardinos2022}. Of these, only one had a predictive performance estimation strategy that could adjusts for multiple model validations (limitedly to SO problems and not affecting model selection), while most of the tools have the necessity to withold a test set for an unbiased estimation of the performance of the winning model, thus loosing samples from the final model training.

No significant efforts have been directed toward improving performance estimation in MO problems, of which SO problems are a special case and which are more relevant and applicable to biomarker discovery in high-dimensional omics data.
Moreover, while there are approaches for SO problems to enhance the performance estimation of a chosen model, these models are still selected based on unadjusted estimations.
Consequently, this does not lead to any improvement in the actual model selection process. On the other hand, strategies that involve subtracting a constant value from model evaluation metrics, used for ranking multiple solutions, fall also short. This approach alone is insufficient to significantly impact the ranking order and thus remains ineffective for enhancing the model selection process. However, if the adjustment takes into account some characteristic of the models, such as the variance in performance during the validation phase or the feature set size, it might change their order. It may be argued that such an adjustment could improve the model selection if the model ranking would use the adjusted estimations. In summary, to the best of our knowledge, no previous work experimented the effectiveness of methods for mitigating the overestimation in MO problems using ML algorithms. Additionally,
no previous work applied the adjustment to the performance estimation during the optimization of the solution set, thus affecting the selection.

\myBigFig{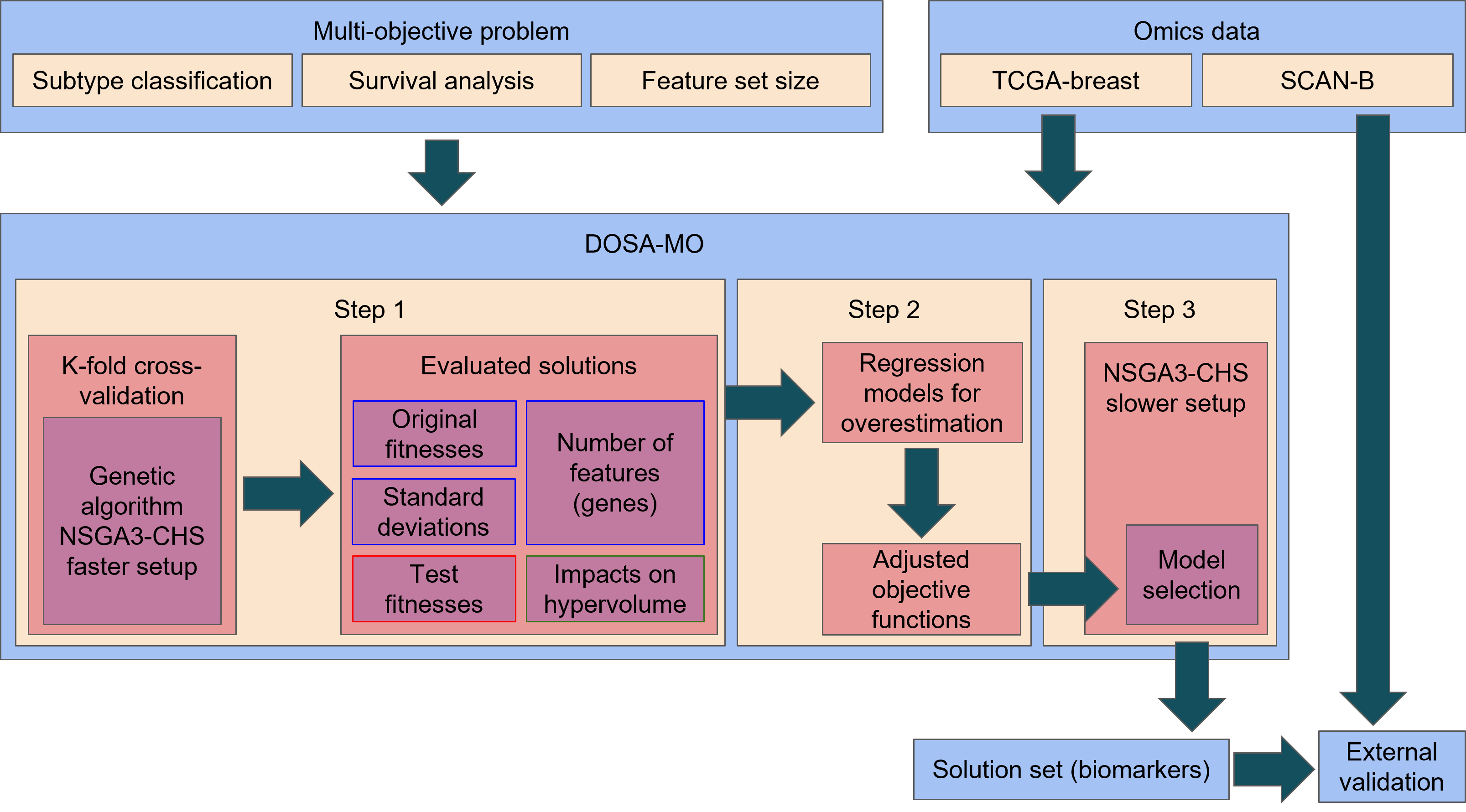}{
Sequence of operations for external validation of DOSA-MO. A MO problem is defined with multiple objectives,
e.g. cancer subtype classification, prediction of survival, and parsimony in the feature set size. A dataset (e.g. TCGA breast omics data)
is fed to DOSA-MO for its optimization process in 3 steps. In step 1 it performs a k-fold CV with a wrapped MO optimizer
(e.g. the NSGA3-CHS GA) and collects the solutions from all the folds. From each solution and objective a sample is constructed.
It has the fitness expected by the wrapped MO, its SD, and the feature set size as independent variables,
the overestimation (expected fitness minus fitness assessed on left-out set) as dependent variable,
and the partial derivative on the HV with respect to this fitness measurement as sample weight.
In step 2 these samples are used to train regression models for overestimation, and new adjusted objective functions are created.
In step 3 a wrapped MO optimizer is run with the adjusted objective functions, impacting how the models are selected.
The solution set is the output of DOSA-MO (e.g. a set of biomarkers).
It might be beneficial to use a faster wrapped optimizer in step 1 than in step 3 since step 1 uses k-fold CV.
An external dataset (e.g. SCAN-B) is used to externally validate the solution set.
}

Here we present the \algorithmName{} (\AlgorithmExtName{}),
a cutting-edge algorithm aimed at predicting and adjusting for overestimation in MO problems.
Initially, a wrapped MO optimizer is integrated with ML algorithms, to create a series of preliminary solutions (feature sets in our case study).
These initial solutions are then used to train regression models that are specifically designed to predict performance overestimation.
This prediction is based on characteristics of the solutions, such as the variance in evaluation metrics (for instance, the variance of balanced accuracy in internal validations) or the size of the feature set itself.
As wrapped MO optimizer for our case study we used a modification of the Non-dominated Sorting Genetic Algorithm III (NSGA3) \cite{Deb2014}:
NSGA3 with Clone-Handling method and Specialized mutation operator (NSGA3-CHS, see \secRef{studyOverview}).
Subsequently, the algorithm executes a wrapped MO optimizer once more, but with a key distinction: it now utilizes the regression models to deliver adjusted fitness evaluations, enhancing the model selection process. The final resulting models are then trained on all the available samples, thus no samples are lost in order to compute their expected performance on new data. In our benchmarking study, which concentrates on selecting gene expression-based feature sets for cancer subtype classification and patient survival prediction,
we have empirically demonstrated that DOSA-MO effectively mitigates overestimation and improves model selection.
It consistently delivers improved performance estimations in biomarker discovery across diverse population-based cohort datasets.
Additionally, two novel measures are introduced in our study: MO Performance Error (MOPE) and Pareto Delta ($P_{\Delta}$).
To the best of our knowledge, these are the first to be designed to evaluate the discrepancy between the performance expected by the algorithm and the actual performance observed on new samples in MO problems.
These assessments consider the entirety of the solution set determined by the MO optimizer, providing a comprehensive view of the algorithm's effectiveness.
The flow of a \algorithmName{} use case is depicted in \figRef{graphical_abstract}.

Disambiguations for the most technical terms can be found in Supplementary Section 3.

\section{Methods}

\subsection{\AlgorithmName{}: algorithm description} \label{sec:adjusted}

The \algorithmName{} algorithm wraps a MO optimizer, serving two purposes: improve the estimate of the solutions' performance, and increase the overall performance of the solution set (set of biomarker models in our case study).
In our case study, these optimizers are represented by genetic algorithms (GAs) \cite{Alhijawi2023} specifically designed for MO problems and paired with supervised ML algorithms,
such as the Gaussian Naïve Bayes (NB) classifier to distinguish cancer subtypes or the Cox Proportional-Hazards Model (Cox) for survival analysis.
In more details, \algorithmName{} consists of three steps.

\textbf{1. Generating solution sets for overestimation prediction.}
	It collects solutions to be used as training samples to learn how to adjust the objective functions that are used to evaluate solutions,
	such as the classification accuracy. This consists of running MO optimizers in a k-fold CV loop.
	For each fold, a solution set is produced using only training data,
	and its performance is measured on the left-out samples.

\textbf{2. Training of regression models for overestimation.}
        For each objective DOSA-MO trains a regression model on the samples collected during step 1.
	Each sample contains as independent variables three meta-features of the solutions that are potentially predictive
	of the overestimation. They are the original fitness (i.e. the fitness used by the optimizer to choose the best solutions,
	measured using only training data, applying inner k-fold CV in our case study), the standard deviation (SD) of
	that fitness (we measure it using bootstrap), and the number of features included in the solution
        (number of genes forming the biomarker in our case study).
	The dependent variable is the overestimation: the difference between the original fitness and the fitness computed on
	new data through CV. Solutions cannot be considered as equally important. We might expect solutions that are in
	crowded areas of the non-dominated front to be selected less often by a decision maker.
        Consequently, each sample is weighted according to
	the partial derivative of the HyperVolume (HV, constructed using the original fitnesses) \cite{guerreiro2021} with respect to the considered solution and objective.
        The weights for each fold and objective are scaled to sum to 1.
        We minimize the absolute error, when allowed by the specific regression model,
	since the impact of an error on the HV is approximately linear for small errors.
	
\textbf{3. Generating the solution set using adjusted performance.}
        A second MO optimizer is deployed to generate a solution set (each solution refers to a feature set in our case study), with objectives 
        that are adjusted by previously trained regression models for overestimation. 
        Each original objective function is replaced by a pipeline that initially computes the function's result, 
        its SD, and the feature count of the solution. 
        This data is then feed to the corresponding adjuster regression model, which predicts the overestimation. 
        The final adjusted performance is calculated by deducting the predicted overestimation from the original fitness value.
        The solution set generated by the MO optimizer during this final run represents the output of the whole \algorithmName{}.

Depending on the MO optimizers used, step 1 might be computationally expensive. A strategy to limit its cost when using GAs is
described in Supplementary Section 1.3.

Step 2 includes computing the SD of the original fitness.
Our method for doing that combining bootstrap with k-fold CV is described in Supplementary Section 1.4.

\subsection{Pseudocode formulation of the \algorithmName{} algorithm} \label{sec:pseudocode}

In order to wrap any MO optimizers, the \algorithmName{} must be polymorphic,
so we define an abstract class \code{MultiObjectiveOptimizer} representing a generic
MO optimizer (\codeRef{moopt}). It has a single method \code{optimize} that works with provided objectives and training data.
The \algorithmName{} is a \code{MultiObjectiveOptimizer} itself (\codeRef{ao}).

\begin{codeFloat}
\begin{smaller}
\begin{verbatim}

class MultiObjectiveOptimizer
   method optimize(objectives, trainingData)
\end{verbatim}
\end{smaller}
\caption{\code{MultiObjectiveOptimizer} abstract class definition.}
\label{code:moopt}
\end{codeFloat}

\begin{codeFloat}
\begin{smaller}
\begin{verbatim}

class DosaMO
      inherits MultiObjectiveOptimizer

   method new(
         tuningOptimizer, adjusterLearner,
         mainOptimizer):
      self.tuningOptimizer = tuningOptimizer
      self.adjusterLearner = adjusterLearner
      self.mainOptimizer = mainOptimizer
	
   method optimize(objectives, trainingData):
      foldsData =
            createFolds(trainingData)
      foldHofs = [
            self.tuningOptimizer.optimize(
                  objectives, f.train)
            for f in foldsData]
      for i in 1::objectives.size:
         obj = objectives[i]
         weights =  [
            assignWeights(
            h.fitnessHyperboxes(), i)
            for h in foldHofs]
         adjuster = trainAdjuster(
               self.adjusterLearner, obj,
               foldHofs, foldsData, weights)
         adjustedObjectives[i] =
               createAdjustedObjective(
                     obj, adjuster)
      return self.mainOptimizer.optimize(
            adjustedObjectives, trainingData)
\end{verbatim}
\end{smaller}
\caption{\code{\AlgorithmCodeName{}} class definition.}
\label{code:ao}
\end{codeFloat}

The method \code{new} is just a simple constructor that saves the polymorphic parts of the algorithm:
the \code{tuningOptimizer}, a MO optimizer used to create the samples for the adjusting regression, the actual
regression model (\code{adjusterLearner}), and the \code{mainOptimizer} that is the MO optimizer that uses the
adjusted objectives to produce the results for the user.

The \code{\AlgorithmCodeName{}} implementation of \code{optimize} first organizes the data into folds.
The resulting object, \code{foldsData}, contains the data itself and also the description of how it is partitioned into folds.
\code{\AlgorithmCodeName{}} then executes the \code{tuningOptimizer} on each fold and collects the results: a set of solutions for each fold.
The set of solutions returned is optimizer-dependent in general, but in our experiments we used
the non-dominated front of all the solutions that were explored.
For each objective \code{obj} the algorithm assigns weights to the solutions: for each \code{tuningOptimizer}
result set \code{h},
the solutions receive a weight that is proportional to the partial derivative of the HV of the belonging
result set with respect to the solution and the current objective \code{obj}.
An adjuster regression model is trained for the current objective \code{obj} using the function \code{trainAdjuster}
that receives in input a regression model \code{adjusterLearner}, the current objective \code{obj},
the \code{tuningOptimizer} solution set for each fold, the data including folds information (\code{foldsData}),
and the weigths of the samples (\code{weights}).
The returned \code{adjuster} predicts how much the fitness of a solution changes between the training sample set
and an unseen testing sample set. The function \code{trainAdjuster} has its own description below.
Using the previous objective \code{obj} and the \code{adjuster}, a new adjusted objective is created that when
evaluating a solution first uses the old fitness function to compute a temporary fitness and its SD,
then adjustes this fitness by subtracting the prediction of the \code{adjuster}.
Finally, \code{\AlgorithmCodeName{}} runs the \code{mainOptimizer} on the whole \code{trainingData},
using the adjusted objectives instead of the original \code{objectives}.

The \code{trainAdjuster} function trains a fitness adjuster regression model for one of the objectives using as samples
the solutions resulting from running the \code{tuningOptimizer} on all the folds defined by \code{foldsData}.
Each solution is assigned a weight proportional to the HV partial derivative with
respect to the considered objective, with the weights for each fold that sum to 1.

\begin{codeFloat}
\begin{smaller}
\begin{verbatim}

trainAdjuster(
      adjusterLearner, obj, foldHofs,
      foldsData, weights):
   for i in 1::foldsData.size:
      allFitnesses = 
            evaluateWithCV(
                   foldHofs[i], obj, foldsData[i])
      originalFitnesses[i] = allFitnesses.innerCV())
      stdDevs[i] = allFitnesses.innerCV_sd()
      testFitnesses[i] = allFitnesses.test()
      nFeatures[i] =
            [h.num_features() for h in foldHofs[i]]
   return adjusterLearner.fit(
         originalFitnesses, stdDevs, nFeatures,
         testFitnesses, weights)
\end{verbatim}
\end{smaller}
\caption{\code{trainAdjuster} function definition.}
\label{code:ta}
\end{codeFloat}

For each fold \code{i} as defined by \code{foldsData}, \code{trainAdjuster} prepares the samples for training the adjuster regressor (\codeRef{ta}).
The samples are prepared separately for each fold, then used together in training.
The function \code{evaluateWithCV} assigns two fitnesses to each solution: one previously computed on the train data of the current fold \code{i}
(using a nested k-fold CV in our case study),
and another computed on the test data. It also computes the SD of the train fitness with the bootstrap method
(Supplementary Section 1.4).
The differences between train and test performance are the values that the regression will learn to
predict. The regression has 3 input meta-features: the original fitness, i.e. the performance on the train data, the SD of the original fitness,
and the number of the features that are included in the solution
(number of genes in our case study).

\subsection{Case study overview} \label{sec:studyOverview}

We have benchmarked \algorithmName{} in the context of MOFS for cancer biomarker discovery.
Our goal was to identify biomarkers for classifying cancer subtypes and survival prediction in kidney and breast cancer patients.
This was done using gene-based transcriptomic datasets from The Cancer Genome Atlas (TCGA) project \cite{hutter2018cancer}.
For breast cancer, an additional cohort from The Sweden Cancerome Analysis Network - Breast (SCAN-B) \cite{Brueffer2018} serves as an external validation set.
A description of the datasets and their preprocessing is in Supplementary Section 1.5.
As wrapped MO optimizer, we used a novel modification of NSGA3 \cite{Deb2014}: NSGA3-CHS, with NB or Support Vector Machine (SVM) as inner classifiers, and Cox as inner survival model.
We used both internal k-fold CV and external validation to compare the unadjusted optimizer (abbreviated as ``zero'') with adjustments by 5 different regression models:
weighted median (dummy), pruned decision tree (ptree), random forest (RFReg), support vector regression (SVR), and SVR with optimized regularization parameters (rSVR).
The regression models are described in detail in Supplementary Section 1.2.

\begin{table*}[h]
\centering
\small
\begin{tabular}{|l|l|l|l|}
\hline
Datasets & Objectives & Inner models  & Validation \\ \hline
Kidney cancer & \shortstack[l]{Subtype classification (balanced accuracy), \\  set size (root-leanness)} & NB  & CV \\ \hline
Kidney cancer & \shortstack[l]{Overall survival (c-index), \\ set size (root-leanness)} & Cox &  CV\\ \hline
Kidney cancer & \shortstack[l]{Subtype classification (balanced accuracy),  \\ overall survival (c-index), \\ set size (root-leanness)} & NB/Cox & CV  \\ \hline
Breast cancer & \shortstack[l]{Subtype classification (balanced accuracy),  \\ set size (root-leanness)} & NB, SVM & CV  \\ \hline
Breast cancer & \shortstack[l]{Subtype classification (balanced accuracy),  \\ set size (root-leanness)} & NB, SVM & External (SCAN-B)  \\ \hline
Breast cancer & \shortstack[l]{Overall survival (c-index), \\ set size (root-leanness)} & Cox &  CV  \\ \hline
\end{tabular}
\caption{Overview of datasets, objectives, inner models, and validation methods used in the analysis of kidney and breast cancer.}
\label{tab:exp_setting}
\end{table*}

NSGA3-CHS is an instantiation of the more general algorithm NSGA*. They are both defined in Supplementary Section 1.1.
The considered experimental setups are listed in \tabRef{exp_setting} and described in detail in Supplementary Section 1.6,
where it is also defined the root-leanness, used as fitness function for the parsimony of the feature sets.
Classification and survival prediction performance are measured respectively with balanced accuracy and concordance index (c-index). 

\subsection{Measuring overestimation in multi-objective problems}

We propose two new methods to measure the error of the performance estimates for the solutions to MO problems: MOPE and $P_{\Delta}$.
The data-driven approach complicates the measurements, as CV yields varied performance metrics. These include the performance anticipated by the optimizer, based on training data, and the performance measured on left-out/new samples post-optimization. To our knowledge, there is no established metric for the error in evaluating the predictive performance of solutions in MO problems within a CV framework.
In SO scenarios, one might simply assess the absolute difference between the fitness expected from training data and the fitness observed on new data. However, in MO scenarios, it is crucial to consider each solution's contribution to the Pareto front. We introduce two novel metrics to measure this estimation error in MO CV setups.

\subsubsection{Multi-objective performance error (MOPE)} \label{sec:methodsPE}

We define the MOPE $E_{\upsilon}$ starting from the HV \cite{guerreiro2021}
computed on the train performance ($H_{\iota}$)
and the Cross HyperVolume (CHV) ($H_{\upsilon}$).
$H_{\iota}$ and $H_{\upsilon}$ have been formally defined by Cattelani et al. \cite{Cattelani2023}. Since $H_{\upsilon}$ is a family of functions,
with the specific instantiation that depends on the user provided function $\upsilon$, $E_{\upsilon}$ is a family of functions too.
\begin{equation}
E_{\upsilon}(X,X')=\abs{H_{\iota}(X,X')-H_{\upsilon}(X,X')}
\end{equation}
Where $X$ encodes the performance of the solutions evaluated on the train data, and $X'$ the performance evaluated on the test data.
In our case study we use the same instantiation of $\upsilon$ as in Cattelani et al.: the function $\lambda$ \cite{Cattelani2023}.

$E_{\upsilon}$ has a simple definition and can be computed very efficiently if the experimental setup already includes the measuring of $H_{\iota}$ and
$H_{\upsilon}$. It can be seen intuitively as the difference between the aggregated performance of all the solution set as expected by the optimizer
taking into account only train data, and the aggregated performance of the same solutions when applied on never before seen test data by decision makers
that must choose their preferred solution informed by train performance only.

\subsubsection{Pareto delta ($P_{\Delta}$)} \label{sec:methodsPD}

The MOPE, derived directly from the training HV and CHV, does not specify the exact sources of discrepancies between these measures. Although CHV accounts for differences between training and testing performance, variations in solution performance of opposite sign can offset each other, resulting in a lower MOPE despite significant train-test discrepancies. To tackle this issue, we introduce a supplementary metric, the $P_{\Delta}$. It has the property of being equal to $0$ only when there is no difference in performance between train and test data for all elements of the solution set. To calculate the $P_{\Delta}$, we sum up the absolute error in fitness estimation for each solution, multiplying it by the derivative of the HV concerning that specific objective and solution. Then, we compute the average across all objectives.

Let $n$ be the number of solutions and $m$ the number of objectives. Let $X$ be an $n\times m$ matrix where
$x_{i,j}$ is the train performance for the $j^{th}$ objective of the $i^{th}$ solution, with $1\le i\le n$  and $1\le j\le m$.
Similarly, let $X'$ be an $n\times m$ matrix where $x'_{i,j}$ is the test performance for the $j^{th}$ objective of the $i^{th}$ solution.
Let $H_\iota$ be the HV computed on the train performance $X$.
$\partial H_\iota / {\partial x_{i,j}}$ is the partial derivative of the HV with respect to $x_{i,j}$.
We define the $P_{\Delta}$ function as
\begin{equation}
P_{\Delta}(X,X')=\frac{1}{m} \sum_{j=1}^{m} \sum_{i=1}^{n}\abs{x_{i,j}-x'_{i,j}}\frac{\partial H_\iota} {\partial x_{i,j}}
\end{equation}
In the special cases with $0$ dimensions or $0$ solutions we define $P_{\Delta}$ as $0$.
As long as the differences $\abs{x_{i,j}-x'_{i,j}}$ are small, $P_{\Delta}$ is proportional to
the difference between the volume of the geometrical union of the space under the train and test fronts and the volume of the intersection of the same two spaces.

\section{Results}

\begin{figure*}[h]
\centering
\begin{tabular}{ccc}
\includegraphics[width=0.3\textwidth]{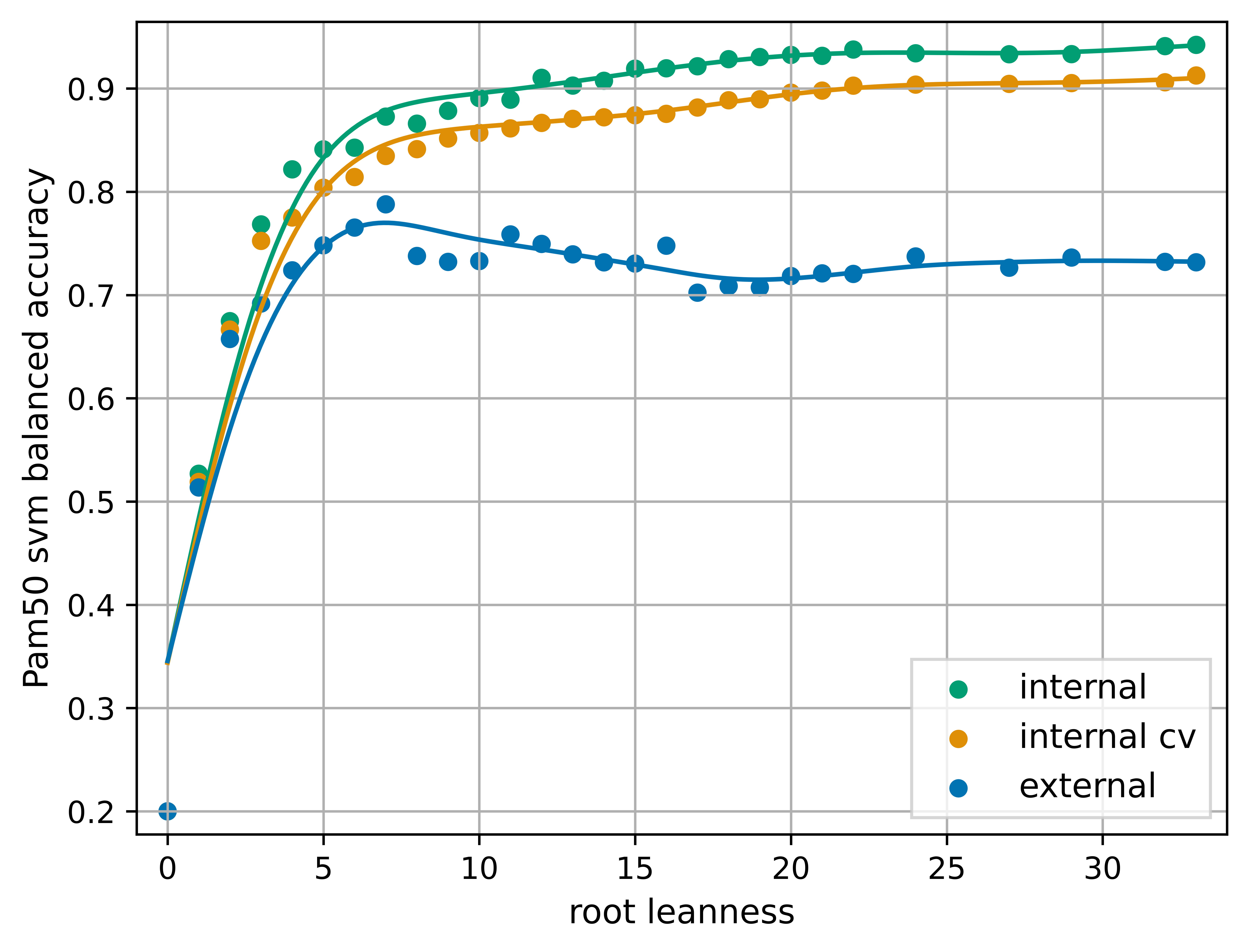} &
\includegraphics[width=0.3\textwidth]{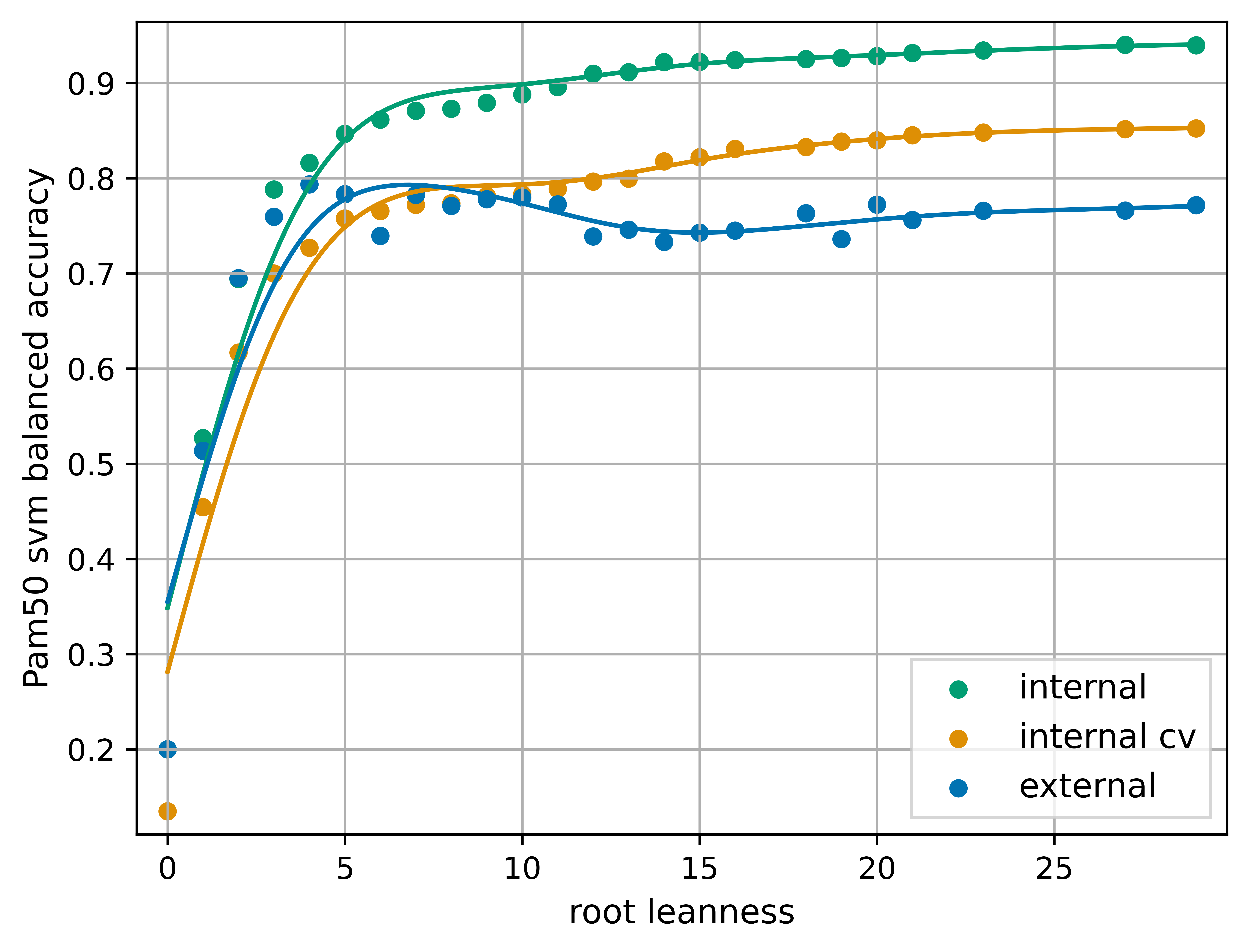} &
\includegraphics[width=0.3\textwidth]{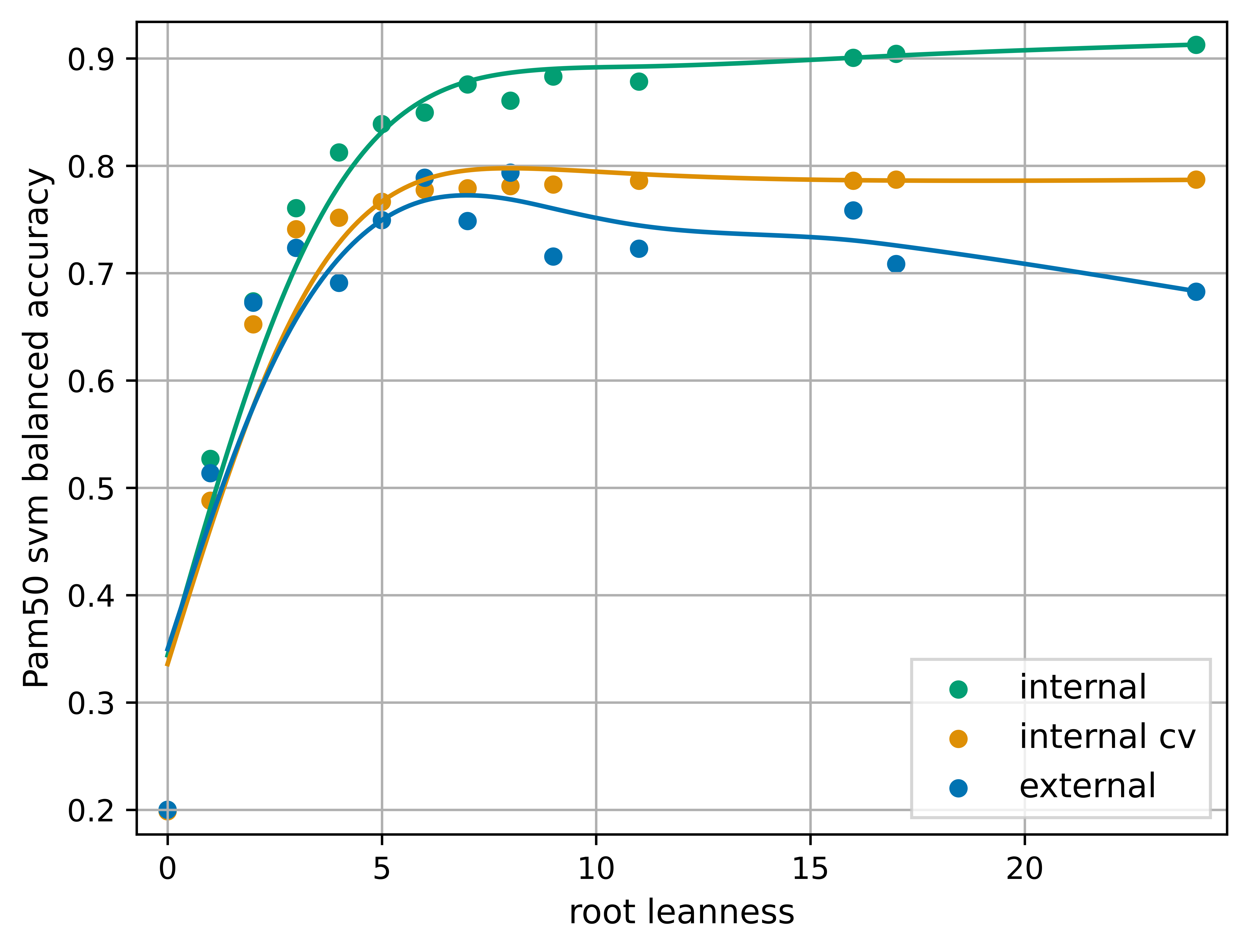} \\
(a) & (b) & (c) \\
\end{tabular}
\caption{
Scatter plots depicting solutions from external validation
on breast cancer transcriptomics data using SVM as inner model.
MO optimization of balanced accuracy for subtypes prediction and root-leanness.
Horizontally, the number of features is depicted for
simplicity. For each solution it is shown the performance
measured in the inner CV, i.e. the performance expected by
the optimizer, the performance of the model trained on the
TCGA breast set and tested on the same set, and the performance
of the same model on the external SCAN-B set. The lines
are interpolating splines.
(a) Using the unadjusted optimizer.
(b) Using SVR as regression model for fitness adjustment.
(c) Using RFReg as regression model for fitness adjustment.}
\label{fig:breast_cancer_scatter_plots}
\end{figure*}

We compare 6 different instantiations of the \algorithmName{} by varying the regression model used for the adjustment,
including the zero regression model, equivalent to not applying any adjustment.
The regression models receive as input 3 meta-features of the solutions correlated
with the overestimation: the original fitness, its SD, and the feature set size (number of genes).
For an example of correlations between meta-features and overestimation see Supplementary Section 2.1.
We use NSGA3-CHS as wrapped model.
The tests are repeated with 8 different combinations of validation type, datasets, objectives,
and classification inner model. For each combination we report two measures for the accuracy of the fitness estimation:
the MOPE (\secRef{pe}) and the $P_{\Delta}$ (\secRef{pd}).
Additionally, we report a measure of overall performance of the optimizers: the CHV (\secRef{chv}).
A focus on the best solution sets from the external validation is shown in \figRef{breast_cancer_scatter_plots}
and discussed in Supplementary Section 2.2.

\subsection{Overestimation in feature selection for biomarker discovery} \label{sec:accuracy}

We measured the error of the performance estimates with two methods: the MOPE and the $P_{\Delta}$.
The MOPE quantifies the discrepancy between the HV computed on the train performance and the CHV.
In contrast, the $P_{\Delta}$ captures differences between expected performance and actual performance on new samples, focusing on the more granular level of individual solutions.

\subsubsection{Comparative analysis using multi-objective performance error} \label{sec:pe}

\begin{figure*}[h]
\centering
\begin{tabular}{ccc}
\includegraphics[width=0.3\textwidth]{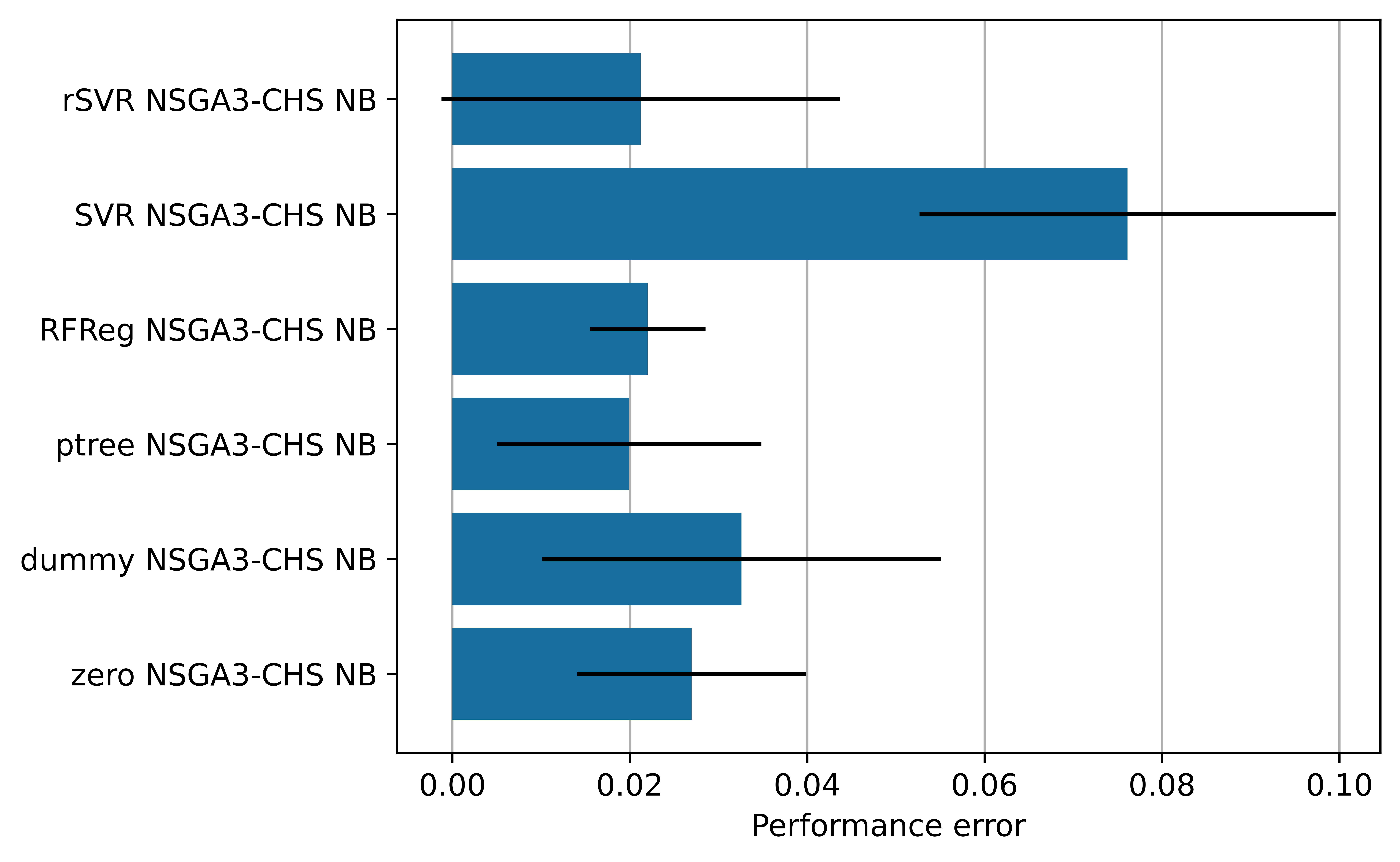} &
\includegraphics[width=0.3\textwidth]{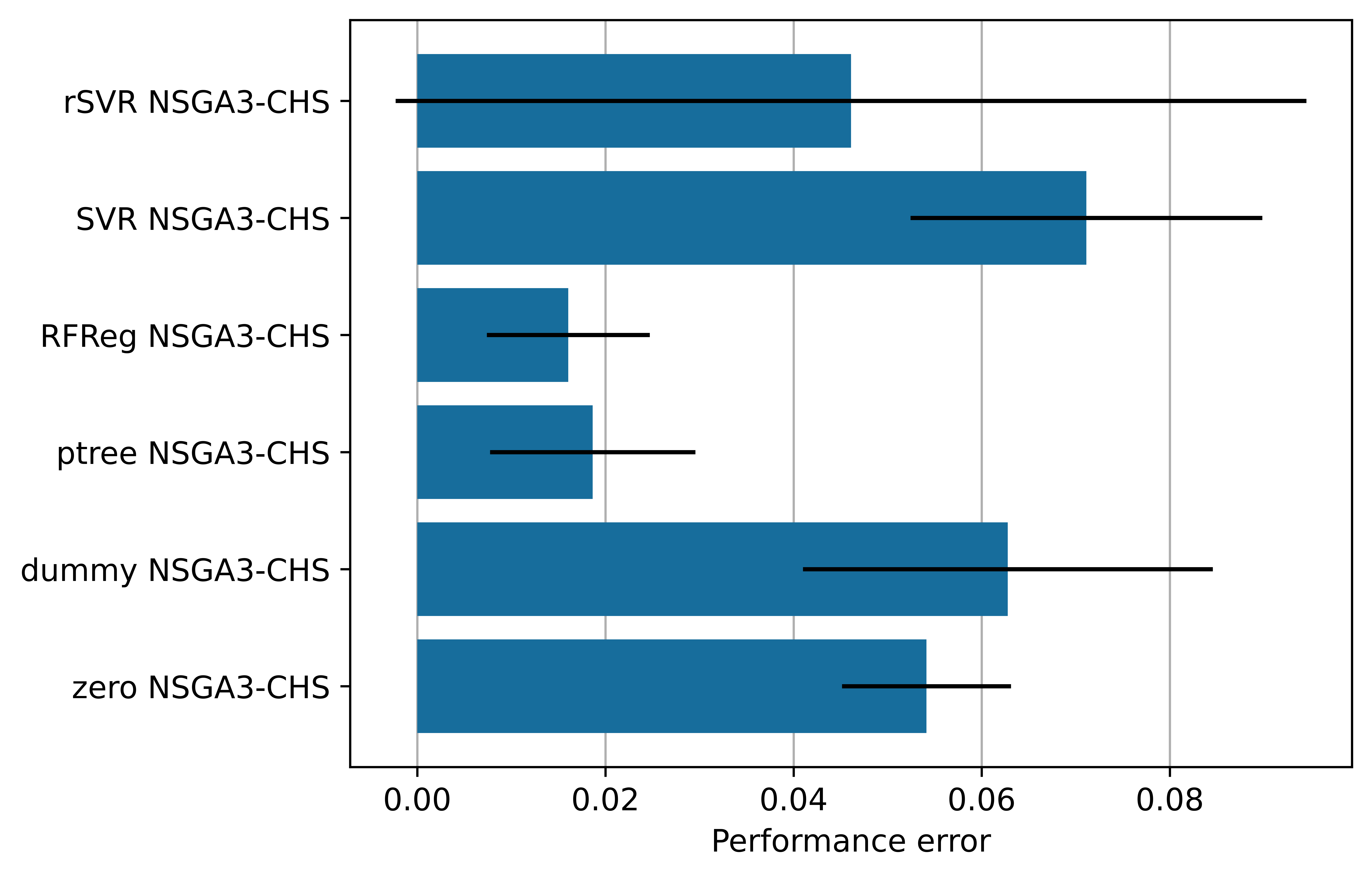} &
\includegraphics[width=0.3\textwidth]{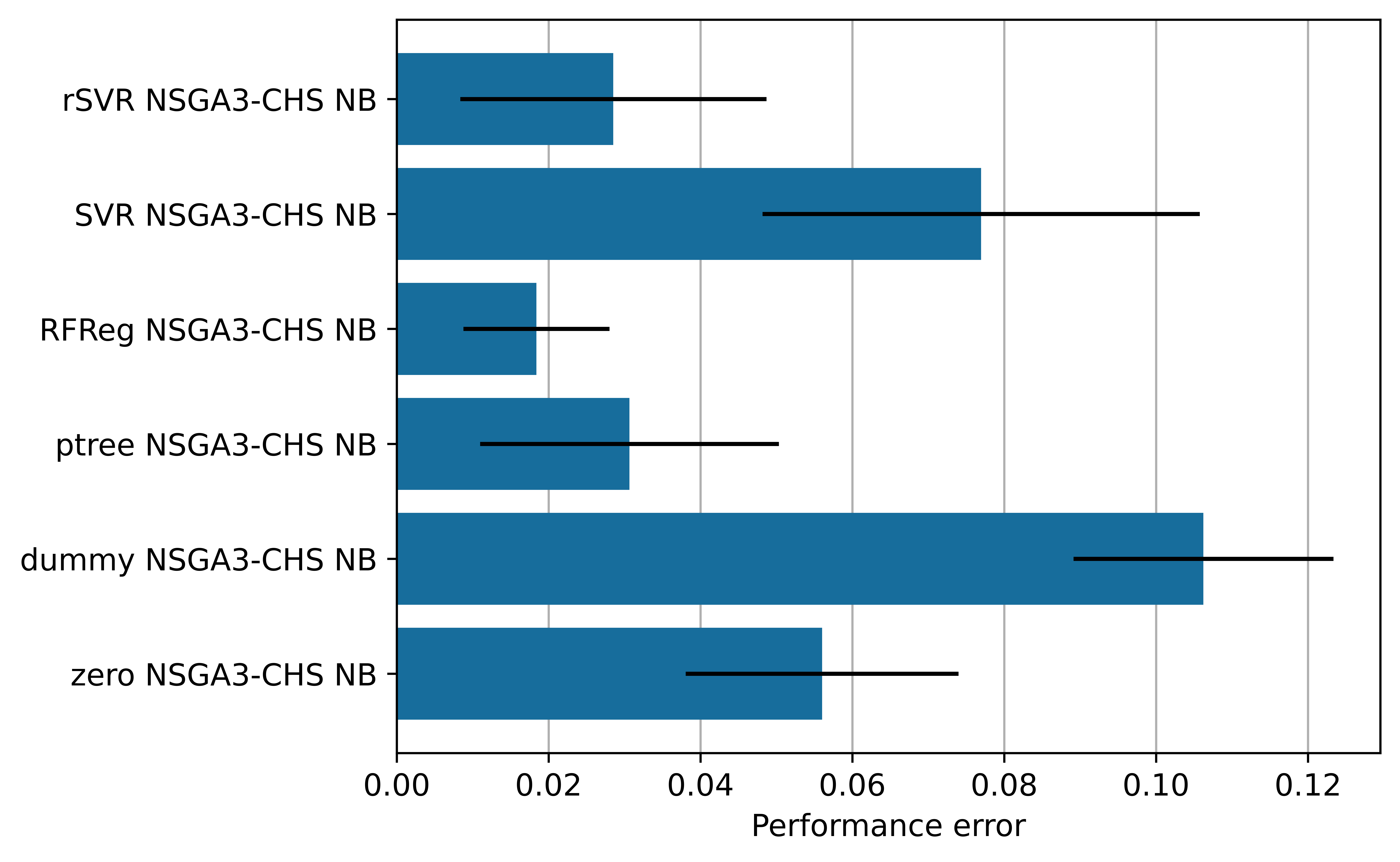} \\
(a) & (b) & (c) \\
\includegraphics[width=0.3\textwidth]{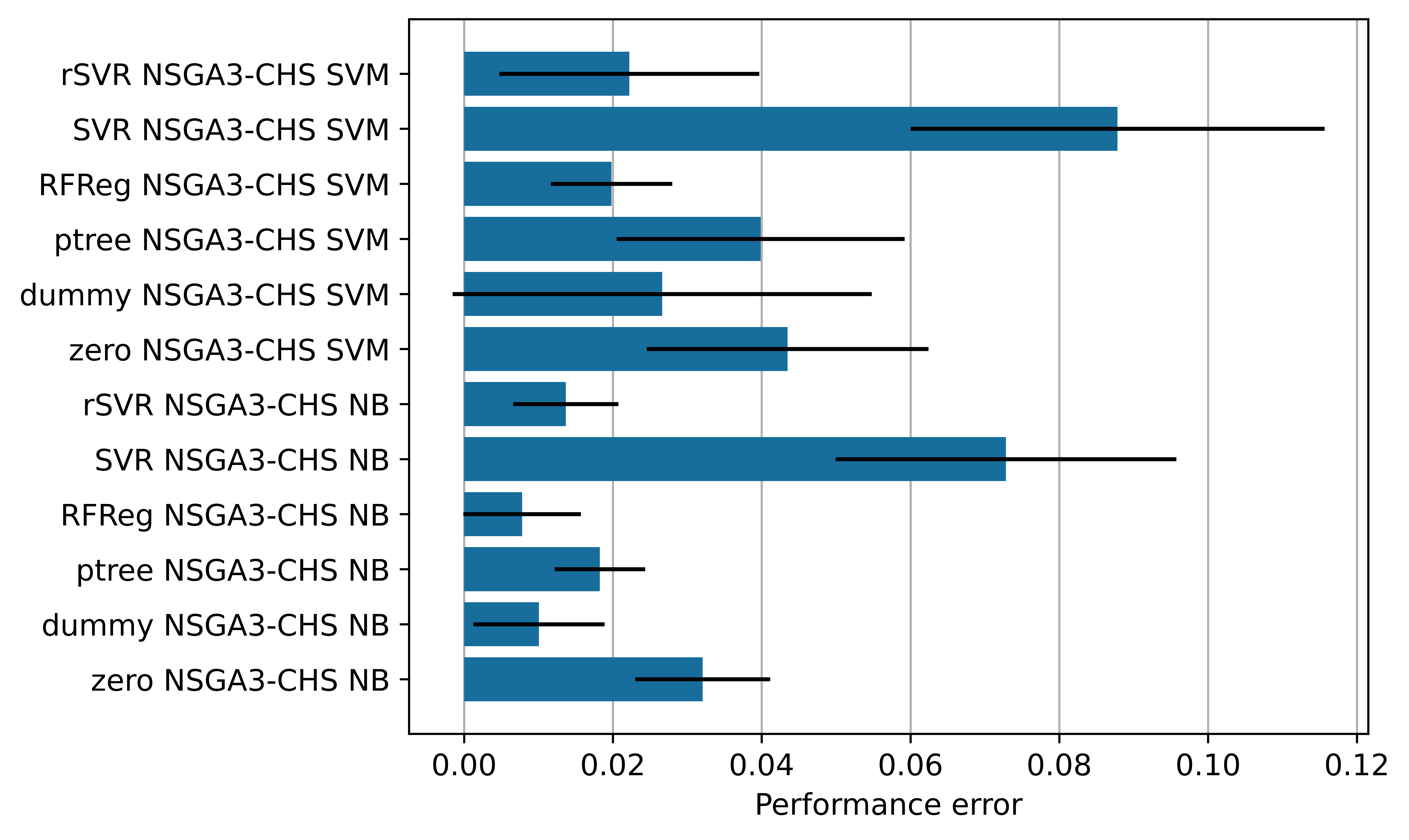} &
\includegraphics[width=0.3\textwidth]{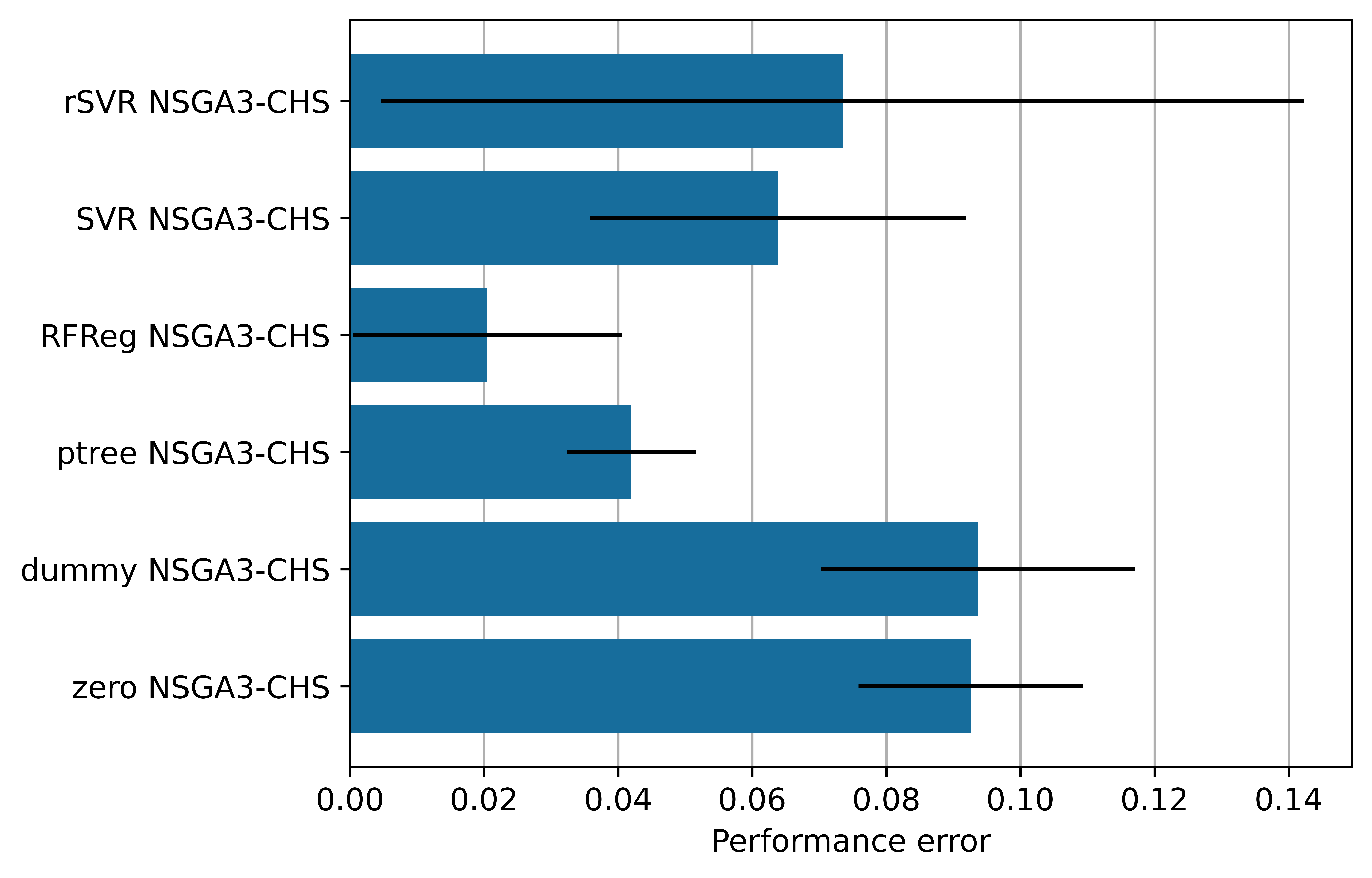} &
\includegraphics[width=0.3\textwidth]{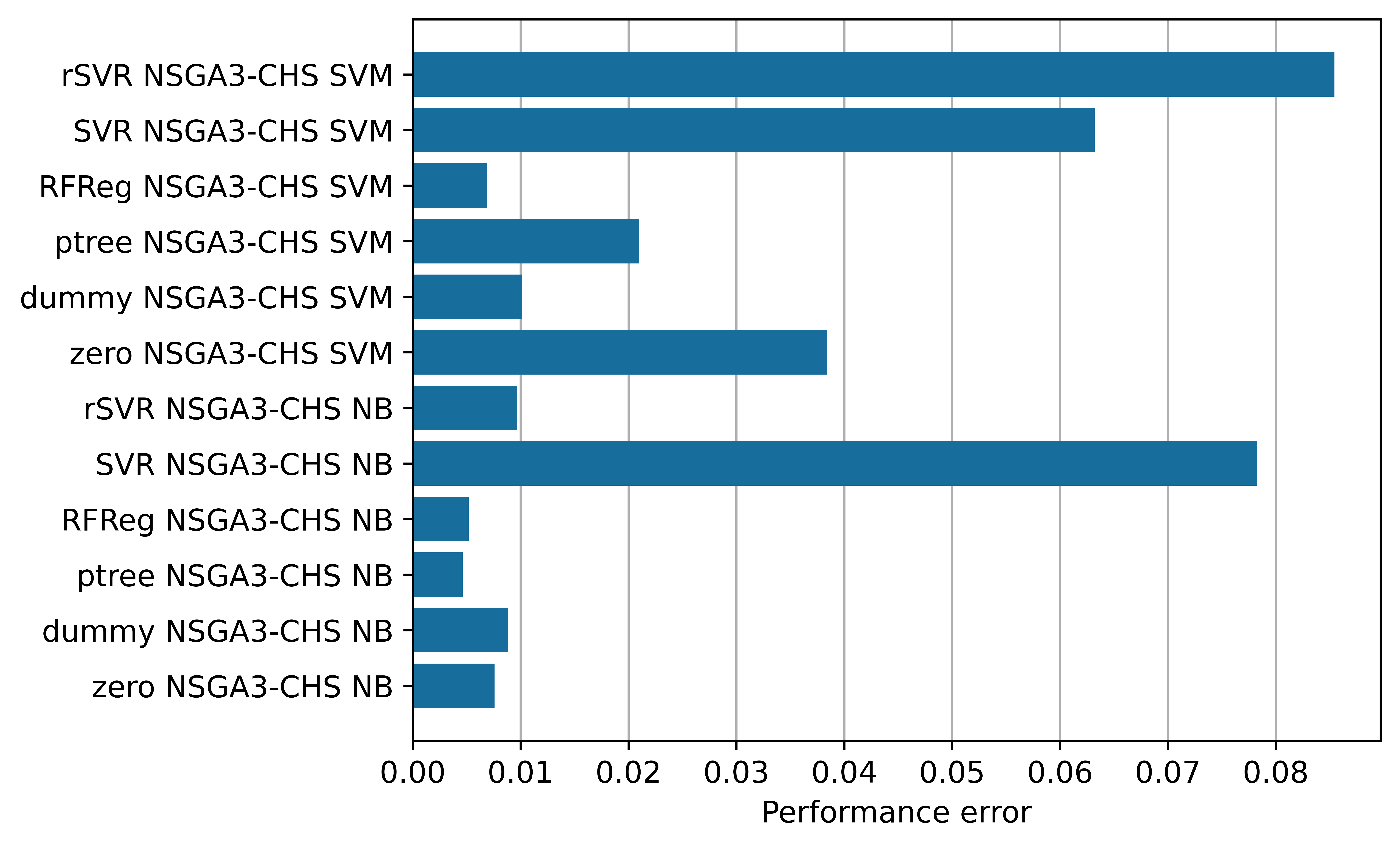} \\
(d) & (e) & (f) \\
\end{tabular}
\caption{MOPE results for internal k-fold CV (a-e) and external validation (f). (a) Kidney cancer, subtype classification and root-leanness. (b) Kidney cancer, overall survival prediction and root-leanness. (c) Kidney cancer, overall survival prediction, subtype classification, and root-leanness. (d) Breast cancer, subtype classification and root-leanness. (e) Breast cancer, overall survival prediction and root-leanness. (f) External validation for breast cancer, subtype classification and root-leanness. Error bars represent SD between folds.}
\label{fig:consolidated_mo_performance_error}
\end{figure*}

The MOPE evaluates the DOSA-MO algorithm's capability 
 in reducing the performance estimation error (\secRef{methodsPE}).
The MOPE outcomes from various regression-based adjusters are reported in \figRef{consolidated_mo_performance_error}.

Significantly, ptree and RFReg often outperform other overestimation predictors. 
Therefore, for users prioritizing the reduction of MOPE, ptree and RFReg regressors emerge as particularly effective choices.
SVR-based estimators of overfitting appear to be the least effective.
ptree and RFReg regressors perform particularly well in external validation scenarios, 
where an external dataset (SCAN-B) is used to evaluate the models (biomarker sets),
 that were selected according to their adjusted performance learned from the initial TCGA breast cohort.

\subsubsection{Comparative analysis using the Pareto delta} \label{sec:pd}

\begin{figure*}[h]
\centering
\begin{tabular}{ccc}
\includegraphics[width=0.3\textwidth]{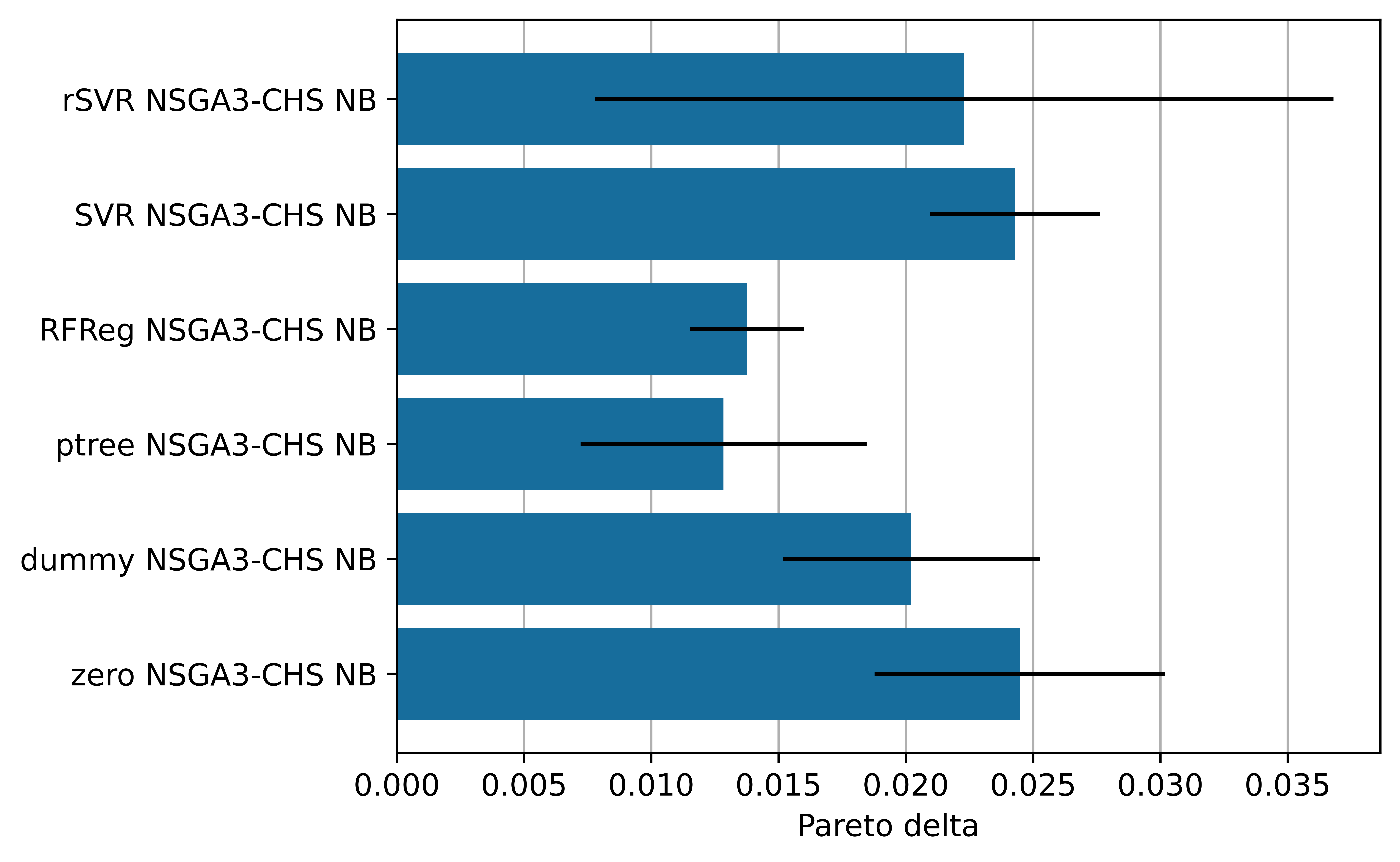} &
\includegraphics[width=0.3\textwidth]{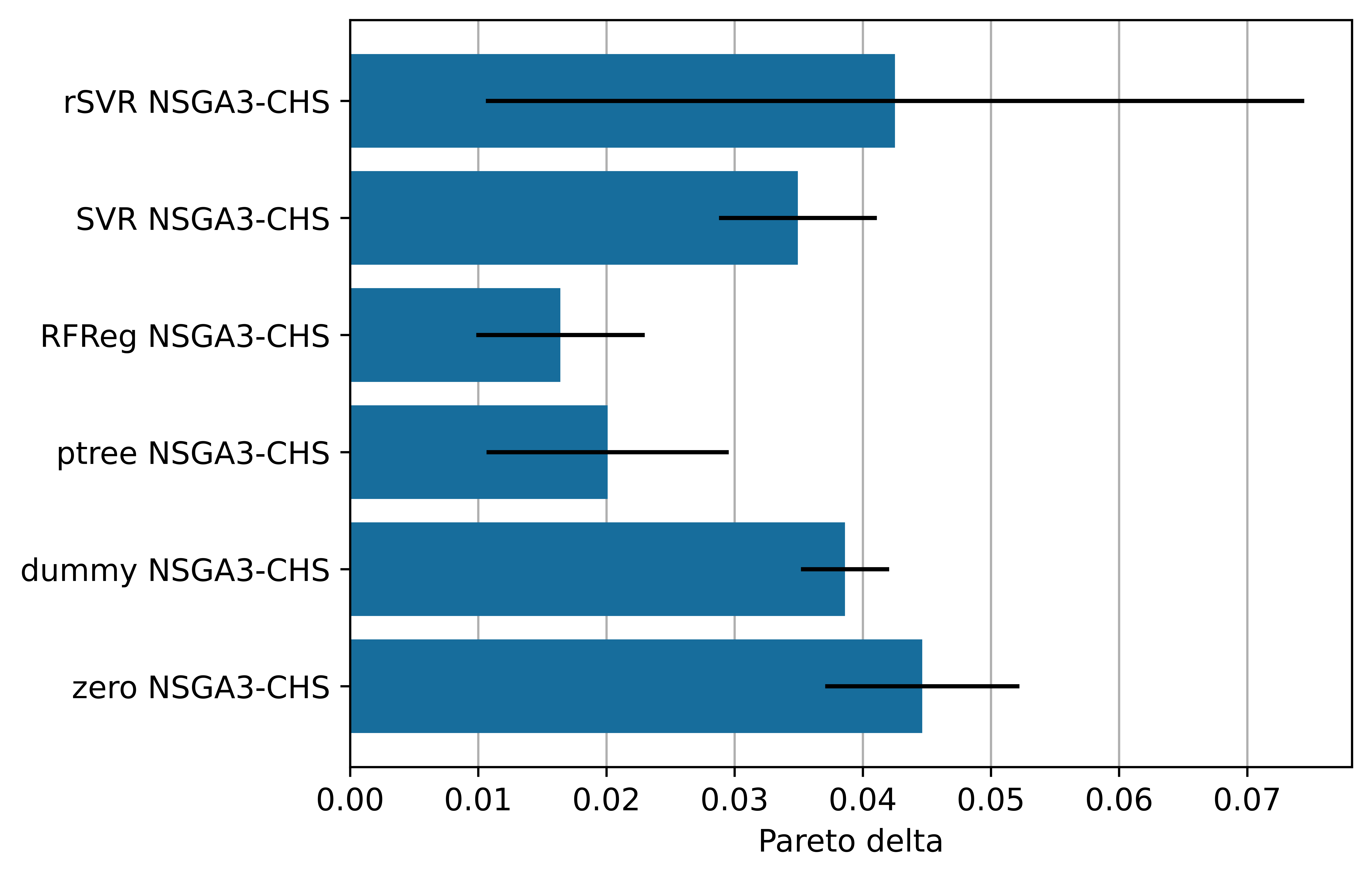} &
\includegraphics[width=0.3\textwidth]{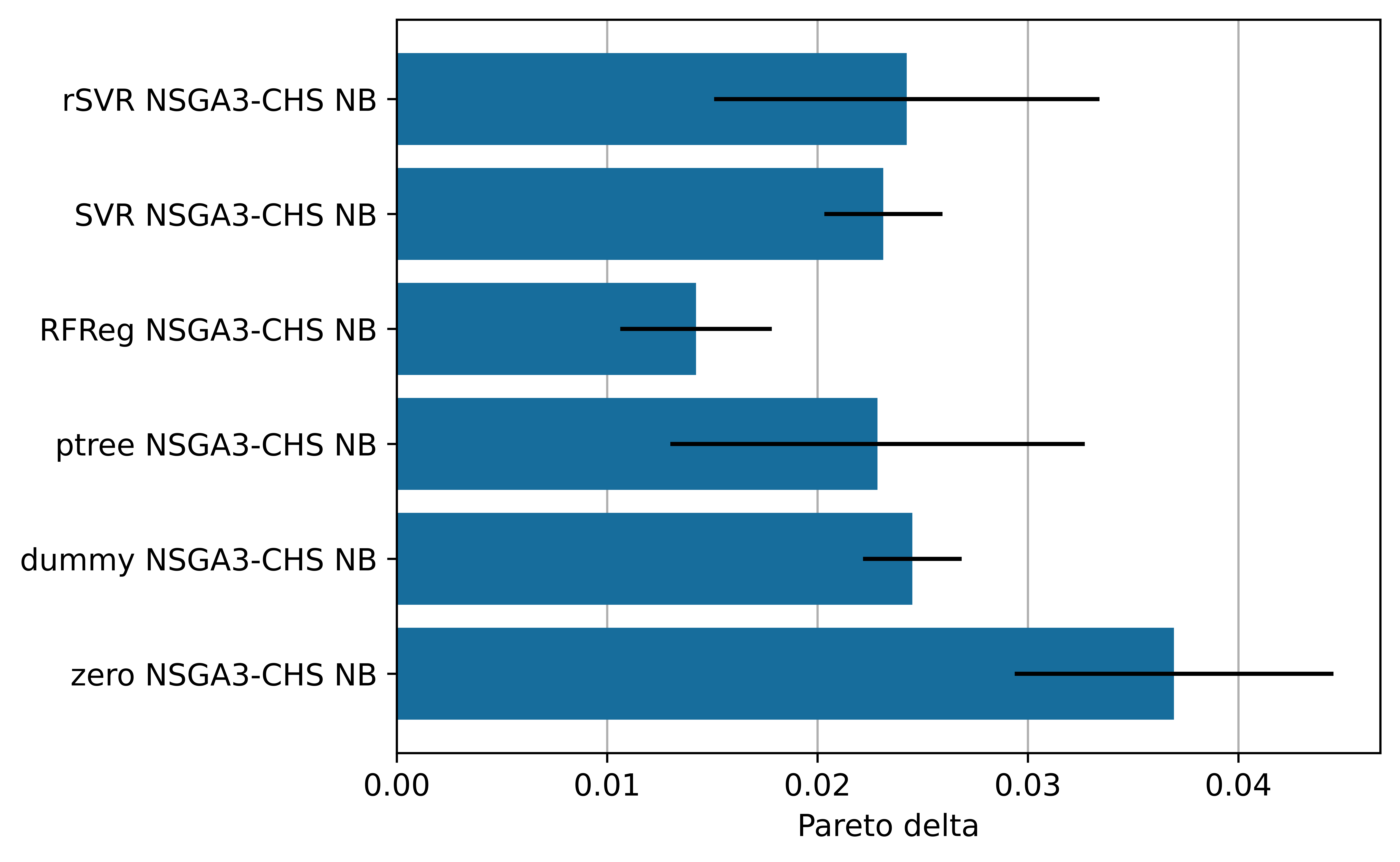} \\
(a) & (b) & (c) \\
\includegraphics[width=0.3\textwidth]{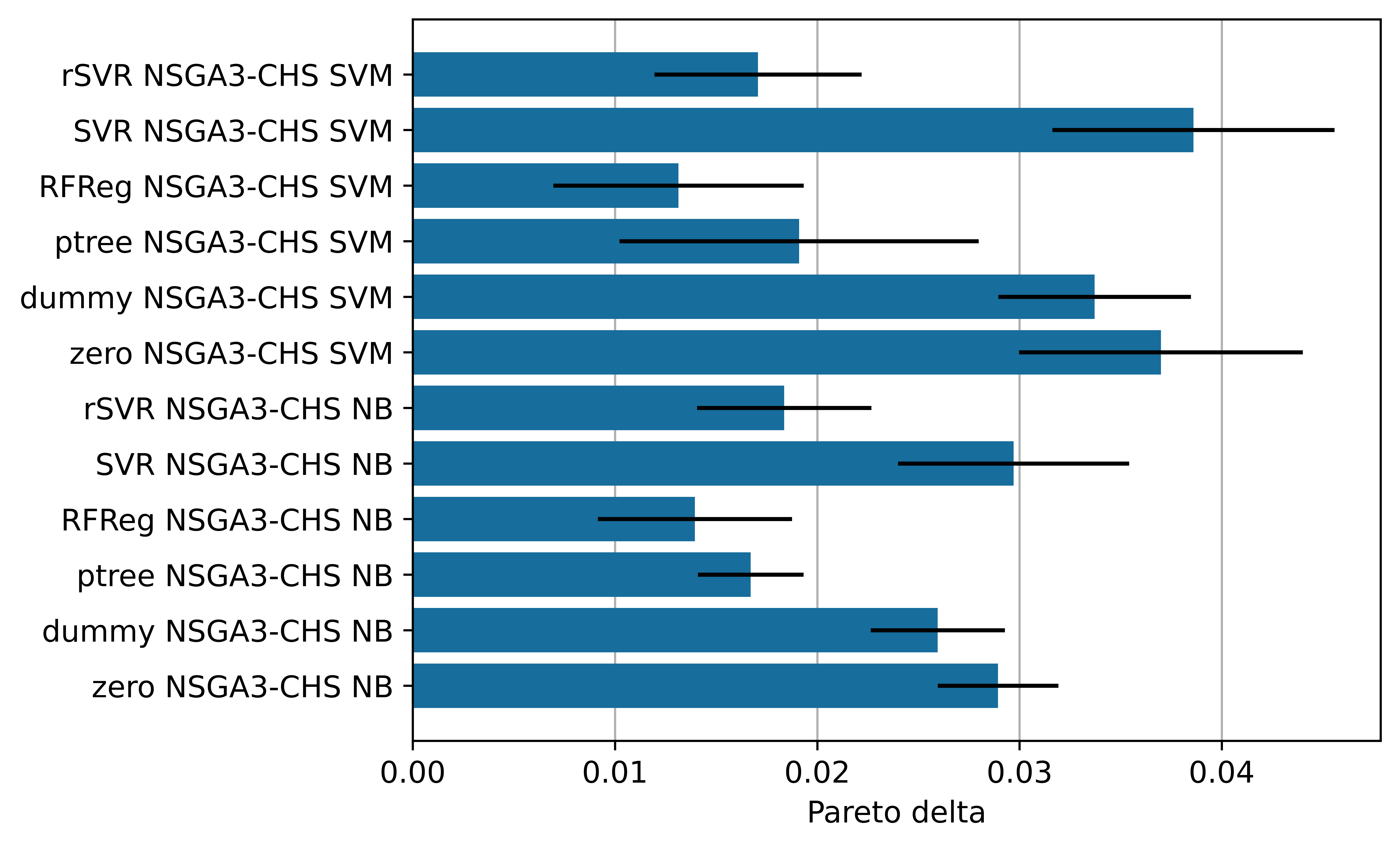} &
\includegraphics[width=0.3\textwidth]{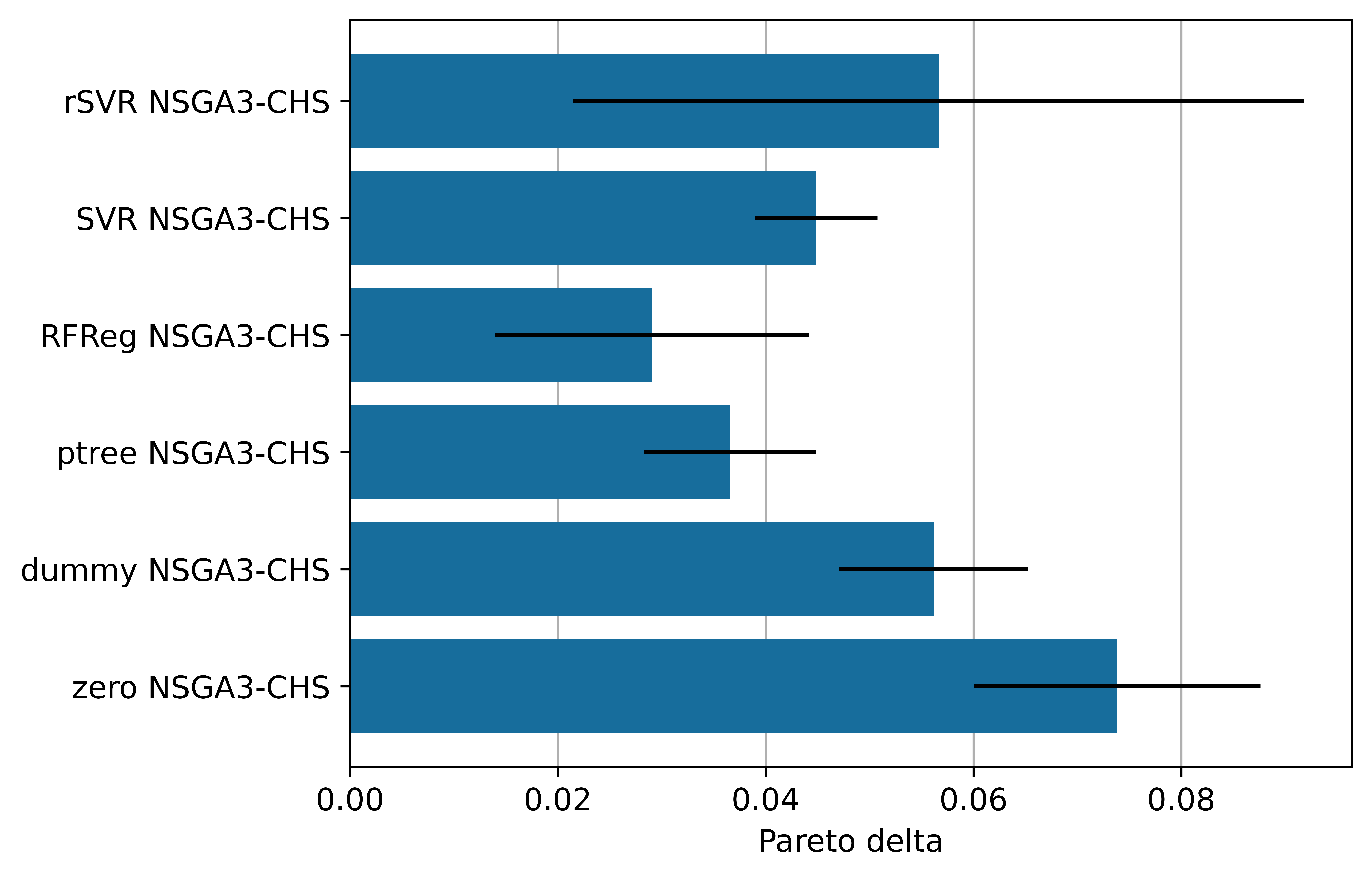} &
\includegraphics[width=0.3\textwidth]{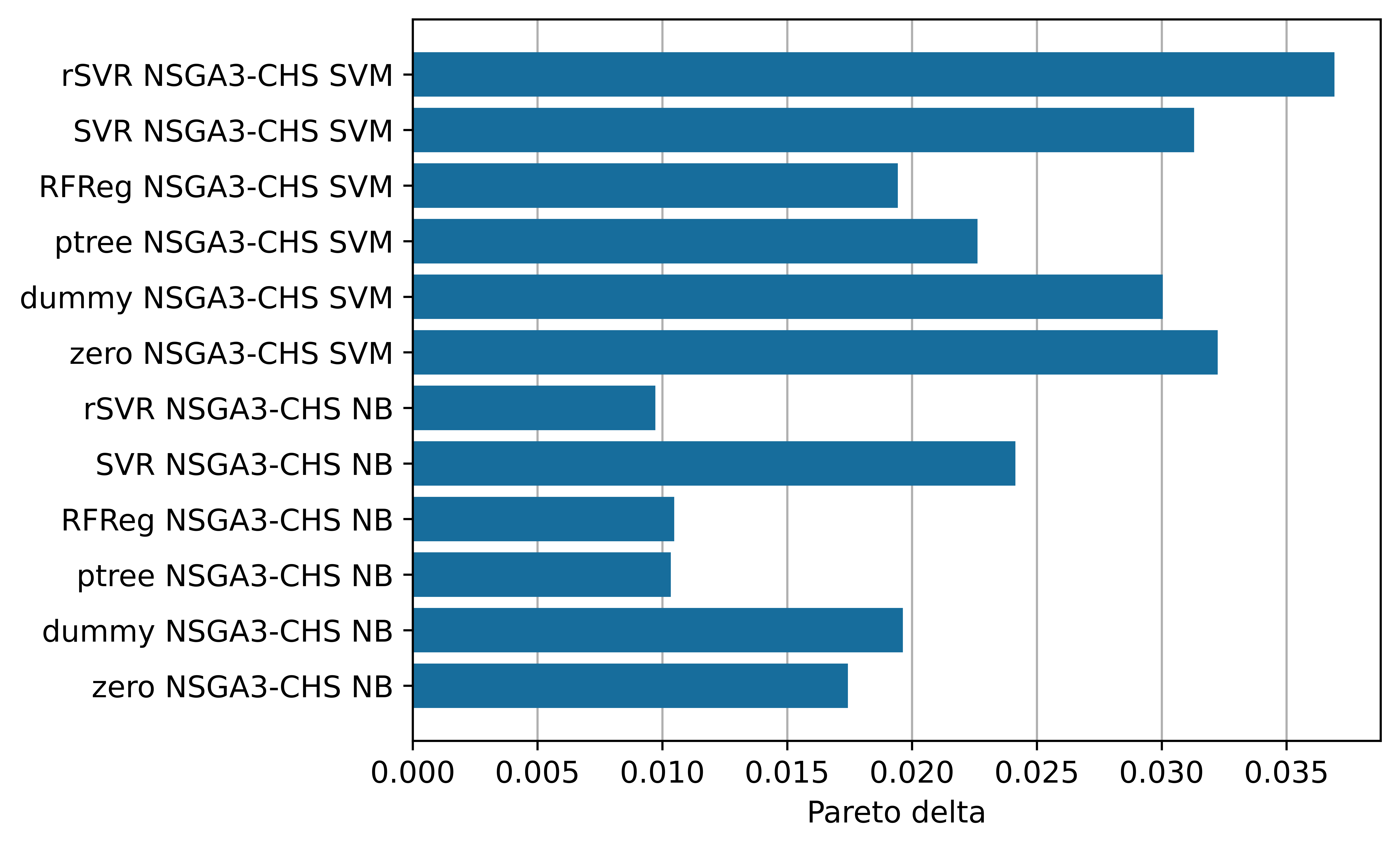} \\
(d) & (e) & (f) \\
\end{tabular}
\caption{$P_{\Delta}$ results for internal k-fold CV (a-e) and external validation (f). (a) Kidney cancer, subtype classification and root-leanness. (b) Kidney cancer, overall survival prediction and root-leanness. (c) Kidney cancer, overall survival prediction, subtype classification and root-leanness. (d) Breast cancer, subtype classification and root-leanness. (e) Breast cancer, overall survival prediction and root-leanness. (f) External validation for breast cancer, subtype classification and root-leanness. Error bars represent SD between folds.}
\label{fig:consolidated_pareto_delta}
\end{figure*}

The $P_{\Delta}$ measure (\secRef{methodsPD}) considers the difference between 
a model's predicted and actual test performance for each individual solution. 
This makes it particularly relevant and more informative than the MOPE in the common situation where
users are presented with multiple solutions but will ultimately select only one.
The $P_{\Delta}$ across the experimental setups is shown in \figRef{consolidated_pareto_delta}.

ptree and RFReg regressors consistently outshine others 
when reducing the $P_{\Delta}$. 
Interestingly, while no significant differences are noted in MOPE between 
the dummy model and the unadjusted optimizer, 
all kidney cancer setups with proper regressors have a lower $P_{\Delta}$ compared to the unadjusted optimizer. 
rSVR performs better than the zero regressor in all scenarios except one.
RFReg and ptree always yield lower $P_{\Delta}$s than the zero regressor, including in breast cancer setups, 
highlighting their promise in both overestimation measures.

\subsection{Evaluating DOSA-MO's effectiveness in MO feature selection} \label{sec:chv}

\begin{figure*}[h]
\centering
\begin{tabular}{ccc}
\includegraphics[width=0.3\textwidth]{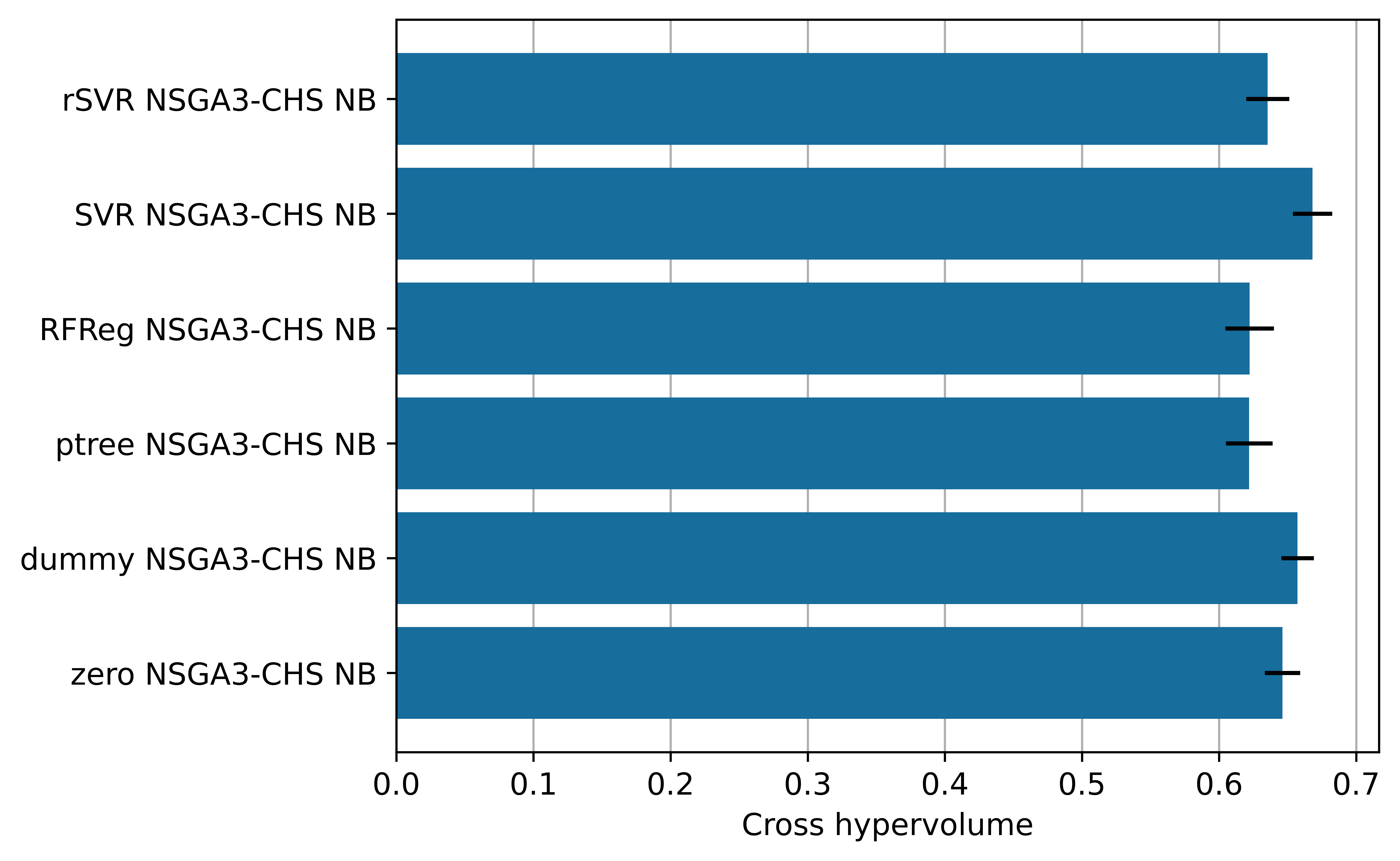} &
\includegraphics[width=0.3\textwidth]{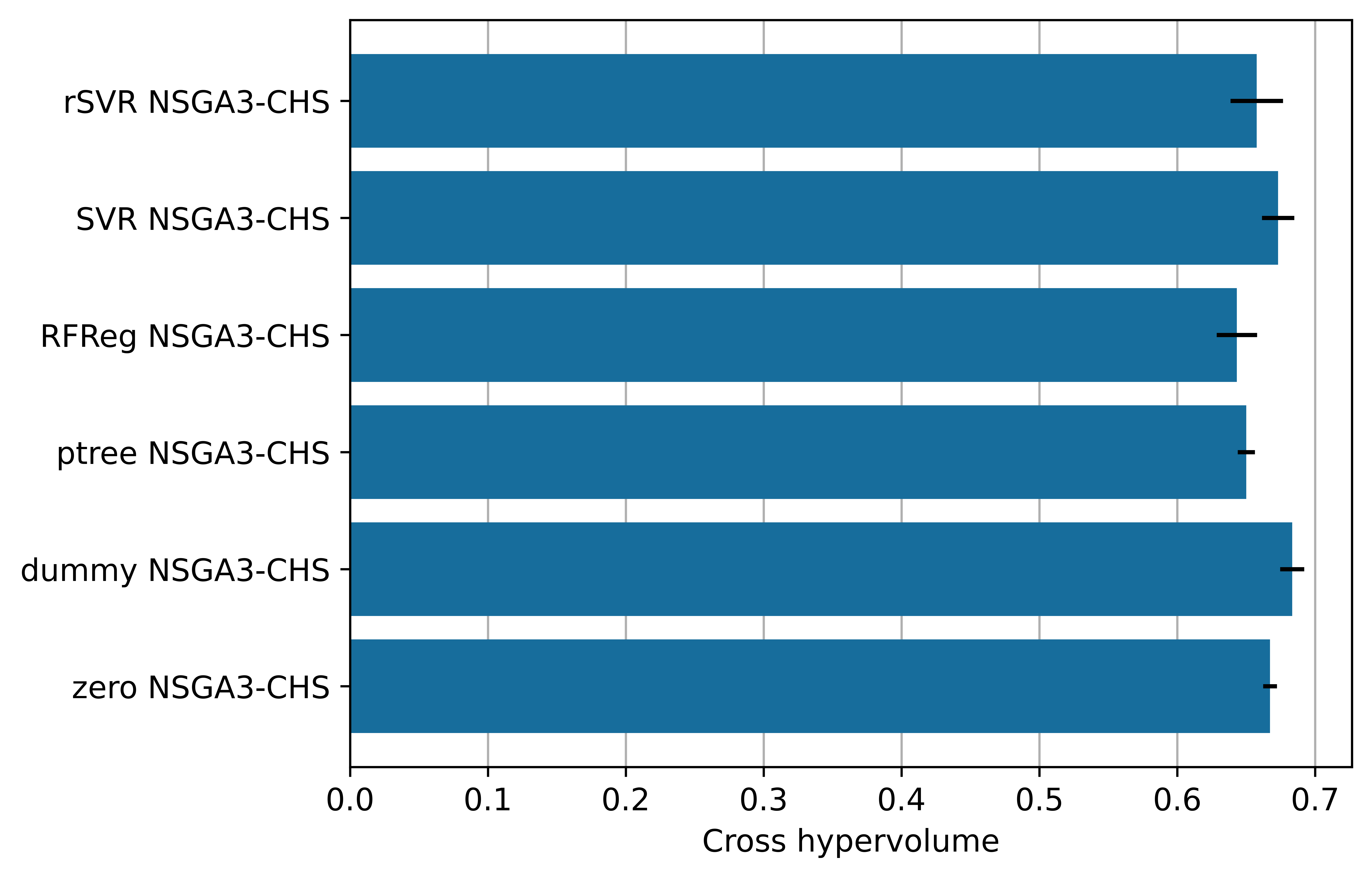} &
\includegraphics[width=0.3\textwidth]{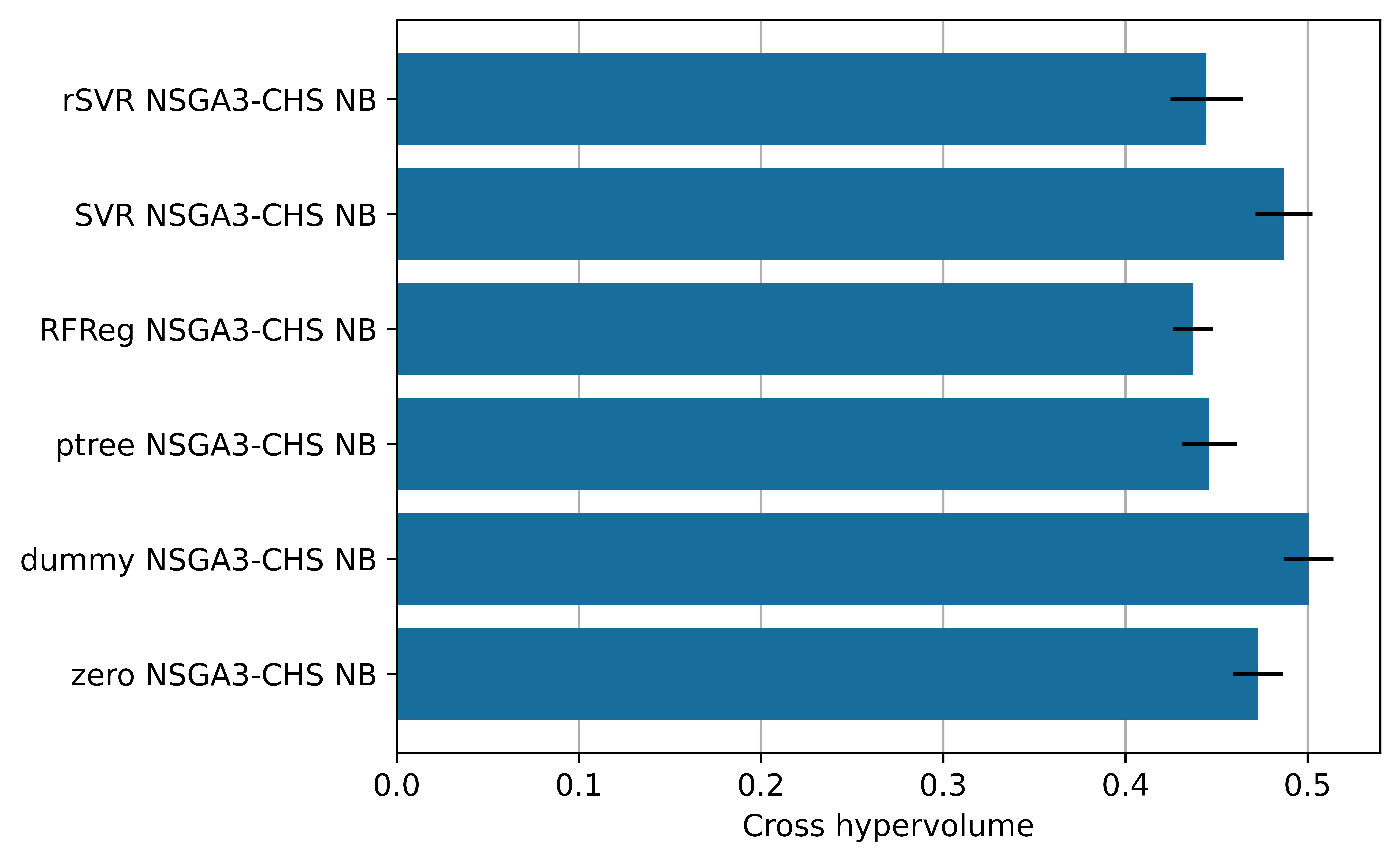} \\
(a) & (b) & (c) \\
\includegraphics[width=0.3\textwidth]{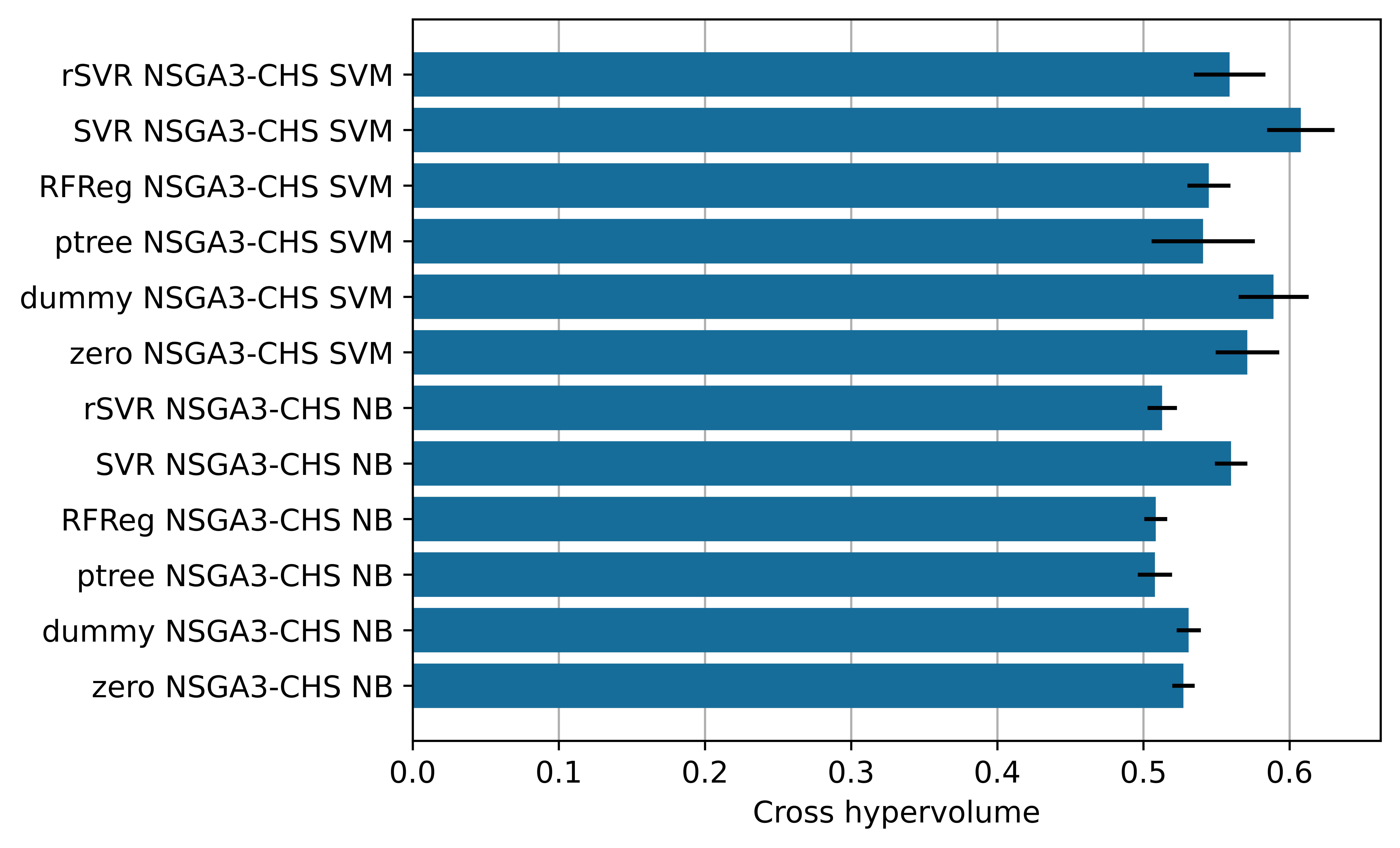} &
\includegraphics[width=0.3\textwidth]{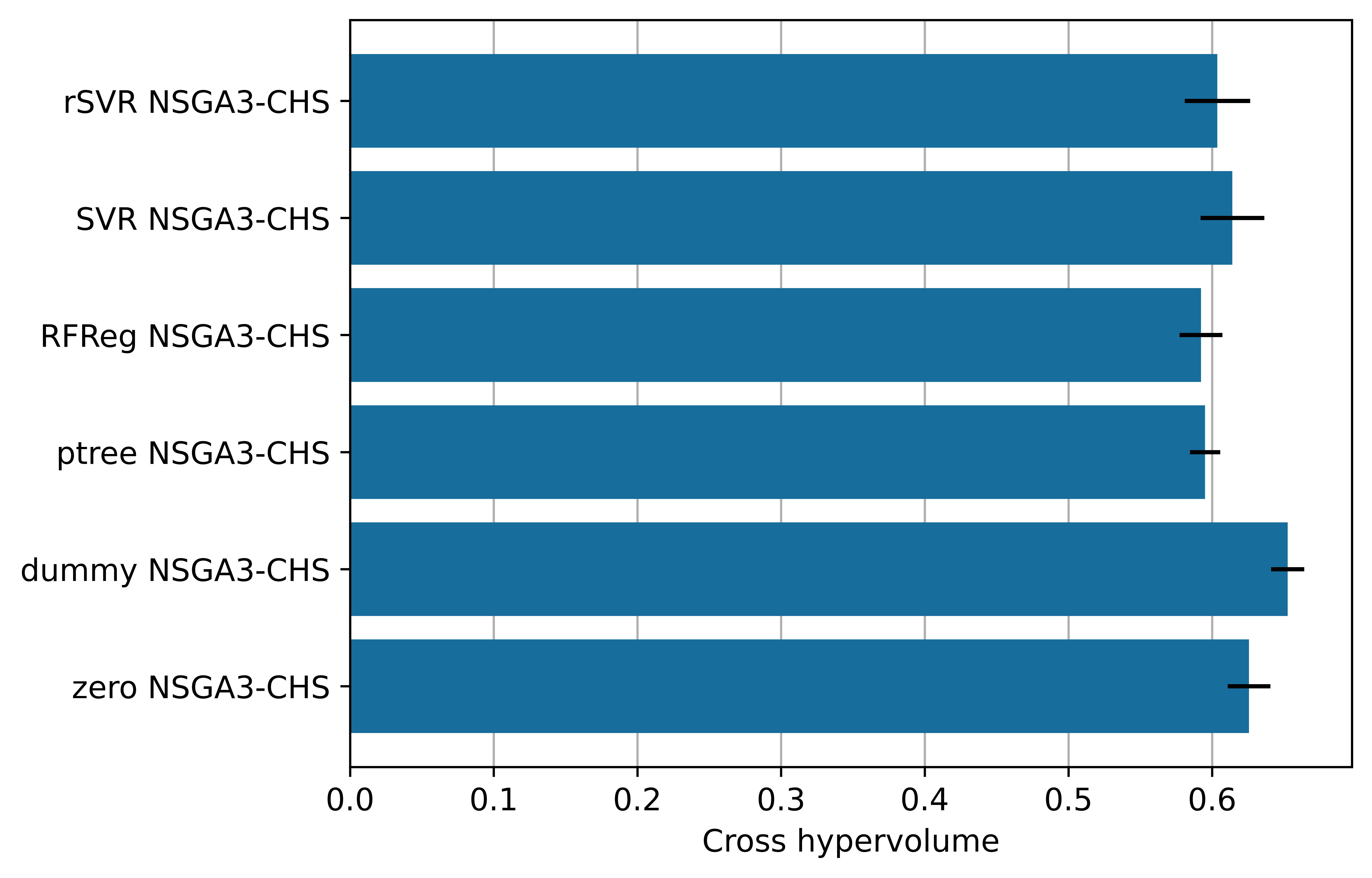} &
\includegraphics[width=0.3\textwidth]{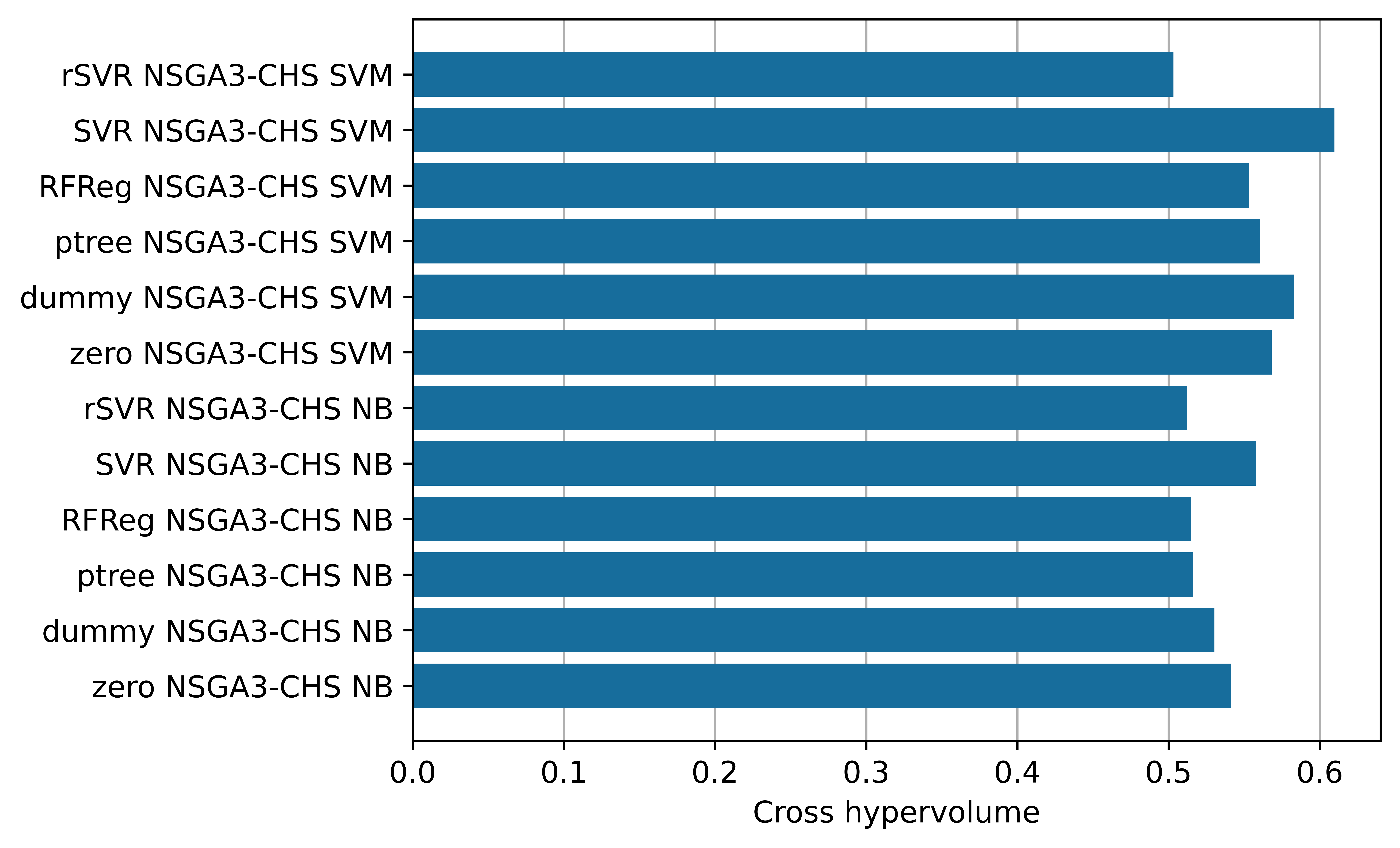} \\
(d) & (e) & (f) \\
\end{tabular}
\caption{CHV results for internal k-fold CV (a-e) and external validation (f). (a) Kidney cancer, subtype classification and root-leanness. (b) Kidney cancer, overall survival prediction and root-leanness. (c) Kidney cancer, overall survival prediction, subtype classification and root-leanness. (d) Breast cancer, subtype classification and root-leanness. (e) Breast cancer, overall survival prediction and root-leanness. (f) External validation for breast cancer, subtype classification and root-leanness. Error bars represent SD between folds.}
\label{fig:consolidated_chv}
\end{figure*}

The previous results highlight DOSA-MO's success in reducing performance estimation errors in ML-driven feature selection for biomarker discovery. 
It is also important to evaluate if DOSA-MO enhances the quality of the produced feature sets:
the final output for biomarker selection models considered for clinical validation. 
To assess the impact on the overall feature selection process, we calculated the CHV for each experimental setup (see \figRef{consolidated_chv}).

Tree-based regressors reduce estimation errors 
but do not significantly improve the quality of solution sets generated by \algorithmName{}. 
This indicates that using these regressors for overestimation prediction may not always enhance solution sets, 
despite reducing quality estimation errors. 
Interestingly, dummy and SVR-based MO optimizers consistently outperform the CHV of unadjusted models, 
highlighting the effectiveness of overestimation prediction for superior feature selection outcomes. 
Specifically, SVR yelds the best CHV in all setups except in scenarios involving the overall survival objective,
while the dummy model excels in setups that include survival prediction.

\section{Discussion}

We introduced DOSA-MO, a novel wrapper MO optimizer designed to adjust performance measures for overestimation in ML-driven MO problems, improving the solution sets that are produced.
This approach has been successfully validated across multiple transcriptomic-based datasets commonly used for biomarker discovery in cancer subtype classification and survival prediction.
Our benchmarking extends to a scenario where biomarker models, trained in one population-based cohort (such as TCGA),
are applied to classify cancer subtype in a second cohort (such as SCAN-B).
This method effectively enables biomarker discovery and validation across multiple cohorts.
Additionally, we have developed two innovative measures for evaluating the performance of MO algorithms: the MOPE and the $P_{\Delta}$.
According to both metrics, the DOSA-MO algorithm demonstrated improved performance estimates in all tested cases,
particularly when using decision tree-based regression models for predicting overestimation.

Our study found that even a basic regression model, which learns solely the weighted average of overestimation and is used for fitness adjustment, resulted in improved overall performance compared to the unadjusted optimizer in 7 out of 8 experimental setups, as indicated by the CHV metric. Likewise, the MO optimizer directed by the SVR model, which predicts overestimations, surpassed the unadjusted optimizer in 7 out of 8 setups, affirming its effectiveness. Notably, the MO optimizer guided by the simpler dummy regression model excelled in the three setups involving survival prediction, while the SVR model proved superior in the five setups focused on subtype classification and feature set size.
The rSVR regression model, that selects the hyperparameters
for the regularization strength of the SVR with random search, utilized for predicting overestimation, underperformed with respect to dummy and SVR regression models.
A regularization that works well while cross-validating on solutions obtained by an unadjusted optimizer, is not as effective when the GA runs with a bigger population,
more generations, and the adjustment is applied to the evaluation of the individuals.

Predicting the overestimation is more complex when the adjustment to the fitness is applied during the optimization as in our approach.
The fitness adjustments alters the optimizer's exploration path, leading to a divergence in the solution distribution from the one used to train the regressors for overestimations,
even within their applicability domain. This creates a ``moving target'' problem, commonly addressed in other contexts, like artificial neural networks, through incremental optimization across multiple epochs. Although executing more computationally intensive MO GAs is possible, further research is essential to enhance our understanding and address the challenges of the applicability domain and the moving target problem in these contexts.

DOSA-MO is versatile and can be applied to any ML problem, including the ones requiring a MO evaluation framework,
making it particularly well-suited for biomarker discovery in high dimensional molecular datasets.
It has been experimentally validated with cancer transcriptomics data, demonstrating its applicability and effectiveness in this domain.

\section*{Funding}
This work was supported by the Academy of Finland [grant agreements 336275, 332510, and 358037], the Jane and Aatos Erkko Foundation [210026], and the Sigrid Jusélius Foundation.

\section*{Acknowledgments}
The computational analyses were performed on servers provided by the UEF Bioinformatics Center, University of Eastern Finland, Finland.

\bibliography{biblio}

\end{document}


\bibliographystyle{unsrtnat}

\maketitle

\twocolumn

\section{Supplementary methods}

\subsection{NSGA* algorithm} \label{sec:nsgaStar}

The \algorithmName{} is a wrapper algorithm that uses an MO optimizer to produce the training data for the regression models and a (possibly different)
MO optimizer that will run with the adjusted objectives to produce the final results for the user. In our case study we used NSGA3-CHS
algorithm as both tuning and main optimizer. The parameters of population size and number of generations were lower for the tuning optimizer
in order to reduce the computational load.
Modified Non-Dominated Sorting Genetic Algorithm II (NSGA2) algorithms NSGA2-CH and NSGA2-CHS
were first introduced in \cite{Cattelani2022} and further validated in \cite{Cattelani2023}.
In this work we analogously modified NSGA3 with the same capabilities of NSGA2-CHS obtaining NSGA3-CHS.
For this purpose we designed a more general algorithm: NSGA*.

\begin{supCodeFloat}
\begin{smaller}
\begin{verbatim}

class Nsga* inherits MultiObjectiveOptimizer
	
   method new(
         nGenerations, popsize, featureImportance,
         sort, tournament, mutation,
         cloneRepurposing):
      self.nGenerations = nGenerations
      self.popsize = popsize
      self.featureImportance = featureImportance
      self.sort = sort
      self.tournament = tournament
      self.mutation = mutation
      self.cloneRepurposing = cloneRepurposing

   method optimize(objectives, trainingData):
      hof = createNewHallOfFame() 
      featImp = self.featureImportance(trainingData)
      pop = createNewRandomIndividuals(
            featImp, self.popsize)
      for 1::self.nGenerations:
         pop = evaluate(pop, objectives)
         pop = self.sort(pop)
         offspring = self.tournament(pop)
         offspring = crossover(offspring)
         offspring = self.mutation(offspring)
         pop = concat(pop, offspring)
         if self.cloneRepurposing:
            pop = ReplaceClonesWithNewIndividuals(
                  pop, featImp)
         pop = evaluate(pop, objectives)
         hof = updateHallOfFame(hof, pop)
         pop = self.sort(pop)
         pop = pop[1..self.popsize]
      return hof
\end{verbatim}
\end{smaller}
\caption{NSGA* class definition.}
\label{supCode:nsgaStar}
\end{supCodeFloat}

\supCodeRef{nsgaStar} describes the algorithm NSGA*, a further generalization of the generalized Nsga2 algorithm presented in \cite{Cattelani2022}.
NSGA* generalizes NSGA2 and NSGA3, together with their modifications NSGA2-CH, NSGA2-CHS, NSGA3-CH, and NSGA3-CHS.
The differences between the algorithms derived from NSGA2 and NSGA3 reside in the tournament and in the sort routines.
NSGA2 uses a binary tournament by rank in the sorted population \cite{Deb2002}.
NSGA3 uses a random tournament where pairs are selected randomly with replacement \cite{Deb2014}.
Both NSGA2 and NSGA3 use a hierarchical sort where individuals are first sorted by their domination front.
The secondary sorting is different: it is based on the crowding distance in NSGA2 \cite{Deb2002} and on reference point niching in NSGA3 \cite{Deb2014}.
The modifications CH and CHS use a primary sorting according to the clone index \cite{Cattelani2022},
and the original sorting of NSGA2 or NSGA3 as secondary sorting.

The class \code{Nsga*} inherits from \code{MultiObjectiveOptimizer}, so it can be used by itself or as wrapped algorithm inside \algorithmName{}.
\code{Nsga*} has a constructor that memorizes the algorithm-specific parameters and strategies:
the number of generations, population size, feature importance function, sorting algorithm,
tournament strategy, mutation operator, and flag for the use of clone repurposing.
Since the \algorithmName{} uses adjusted objective functions, NSGA* takes an \code{objectives} object in input.
It is used by the \code{evaluate} function and represents the NSGA* ability to run with different objective functions.
The algorithm is similar to generalized NSGA2 \cite{Cattelani2022}, but with the added possibility of personalizing the tournament strategy, sorting algorithm
and objectives. It can be used as inner MO optimizer by the \algorithmName{} and can be specialized also as NSGA3, NSGA3-CH, and NSGA3-CHS.

\subsection{Considered adjusting regression models} \label{sec:regressors}

Five regression models have been considered in the experimental validation for adjusting the fitness functions. All of them use the sample weights,
computed by the \algorithmName{} as the partial derivatives of the HV as explained above.
\begin{description}
  \item[Dummy.]
        The simplest regression model learns the weighted median.
  \item[ptree.]
        The pruned decision tree regression model, minimizing the weighted absolute error. The tree is pruned with
	the Minimal Cost-Complexity Pruning algorithm \cite{breiman1984}. The complexity parameter for the pruning is optimized by running a 5-fold
	CV on its training data.
  \item[RFReg.]
        The random forest regression, minimizing the  weighted absolute error.
  \item[SVR.]
        The epsilon-support vector regression with Gaussian kernel type \cite{Drucker1996}. It uses an l2 regularization penalty.
	We use the default parameters of $C=1$ and $\epsilon{}=0.1$.
  \item[rSVR.]
        The randomized SVR uses random search with 5-fold CV on its training data to optimize the parameters $C$ and $\epsilon{}$.
\end{description}

We include in the tests the unadjusted NSGA3-CHS, that is equivalent to use regression models that predict always 0
(shortened as ``zero'').

\subsection{Limit the computational overhead for adjustments} \label{sec:overhead}

Running the tuning optimizer in a nested k-fold CV in order to generate the samples used for training the regression models
imposes a computational overhead with respect to the cost of running the main optimizer without any adjustment to the fitness functions.
An high overhead could make the \algorithmName{} impractical in real cases. In our experimental validation, that uses GAs as inner MO optimizers,
we limited the computational overhead by using in the tuning optimizer a smaller population and less generations than in the main optimizer.

The simplifying assumption is made for the computational cost of the GAs to be proportional to the number of individual evaluations
multiplied by the number of samples used for evaluating the individuals (in training algorithms the cost is typically at least linear in the number of samples,
since the algorithm has to at least iterate through them),
and for the number of evaluations to be in turn proportional to the population size multiplied by the number of generations
(this is in fact an upper bound considering that individuals equal to previously evaluated ones might not need to be evaluated again).
We define a parameter $\mu{}$ 
as the desired ratio between the computational cost of the tuning phase with respect to the main optimization phase.

In order to have a computation time of the \algorithmName{}
approximately double than the time required by the unadjusted optimizer,  $\mu{}$ is set to 1 in our experiments.

In the case of internal k-fold CV, let $\eta{}$ be the number of external folds. We use the same number of folds also for the generation
of the training samples for the regression models, so the tuning optimizer is executed that number of times for each execution of the \algorithmName{}.
Let $\rho{}$ be the population size for the main optimizer, and $\rho{}'$ the population size for the tuning optimizer. We compute $\rho{}'$ with
the following equation.

\begin{equation}
\rho{}'=\round{\rho{}\sqrt{\frac{\mu{}}{\eta{}-1}}}
\end{equation}

Where $\round{\cdot{}}$ is the round to the nearest integer operation.

Let  $\gamma{}$ be the number of generations for the main optimizer, and $\gamma{}'$ the number of generations for the tuning optimizer.
Similarly, we compute $\gamma{}'$ with the following equation.

\begin{equation}
\gamma{}'=\round{\gamma{}\sqrt{\frac{\mu{}}{\eta{}-1}}}
\end{equation}

According to the previous assumptions, the cost of running the main optimizer $c$ is $c=\rho{}\gamma{}m$, with $m$ being the number of samples.
The cost of running all the iterations of the tuning optimizer $c'$ is the following.

\begin{equation}
c'=\eta{}\rho{}'\gamma{}'\frac{m(\eta{}-1)}{\eta{}}=\round{\rho{}\sqrt{\frac{\mu{}}{\eta{}-1}}}\round{\gamma{}\sqrt{\frac{\mu{}}{\eta{}-1}}}m(\eta{}-1)
\end{equation}

Ignoring the round operations that have just a small contribution it is easy to verify the desired ratio.

\begin{equation}
c'\approx\rho{}\gamma{}\mu{}m=\mu{}c
\end{equation}

So the computational cost of using the adjusted optimizer wrapper, including the main optimizer,
is approximately equal to $(1+\mu{})c$, or even lower if the cost of evaluating individuals grows more than linearly in the
number of samples.

The external validation is faster since it does not have an outer k-fold CV.
For the external validations, we have arbitrarily set $\eta{}=5$, resulting in samples for regression training to be
acquired from 5 folds.

\subsection{Computing the fitness functions variance} \label{sec:variance}

One of the meta-features of the solutions, used for prediction by the adjustment regression models, is the SD of the original fitness.

Collecting the fitnesses of an individual (feature set in our case study) on the different folds
and computing the SD of a performance metric (for example, the balanced accuracy or the \cIndex{}),
would have a limited precision because of the number of the folds, that is necessarily contained to have an acceptable computational time. On the other hand, increasing the fold count would heighten computational demands, as the inner models need to be trained for each fold.
Additionally, each fold fitnesses would be estimated on a smaller left-out sample set.
Moreover, performing repeated CV would increase the number of evaluations but the same test samples would be reused.

Our method to calculate each solution's fitness variances addresses these issues by performing bootstrap analysis within each fold of the k-fold CV, followed by aggregation of the results across all folds. The following steps describe how one of the fitnesses for one of the individuals is computed.

\begin{enumerate}
  \item For each of the folds there are training and testing samples. The inner model is optimized on the training data.
	  The fitness is computed on the whole testing data, then by using bootstrap on the testing data, the variance
	  for the fitness is computed.
  \item The results from each of the folds are aggregated considering the folds as strata in a stratified bootstrap.
          The variances in the different folds are assumed uncorrelated and combined with the equation for the variance of
	  the mean of uncorrelated random variables: the variance of the mean is the sum of the variances divided by the square
	  of the number of folds.
\end{enumerate}

When cross-validating, there are two sources of performance variance: the composition of the training set affects the training process thus can lead to different predictors,
and indirectly to different expected performance, while the composition of the test set directly impacts the expected performance \cite{Rodriguez2010}.
It is known that there is no unbiased estimator of the variance of k-fold CV \cite{bengio2005bias}.
The described procedure accounts for the variance explained by the limited number of test samples, but only partly for the variance explained
by the limited number of training samples, since the training samples are reused in different folds
(this is a well known unavoidable limitation in k-fold CV \cite{Arlot2010}) and the bootstrap is applied on the test sets but not on the training sets,
to avoid incurring in infeasible computational costs. Despite the limitations, this estimation of the fitness variance appeared predictive
of the overestimation in our experiments, thus justifying its inclusion in the meta-features for the overestimation prediction.

\subsection{Benchmarking datasets: description and pre-processing} \label{sec:datasets}

We conducted benchmark studies with two primary goals: firstly, to identify gene expression-based biomarker sets for classifying cancer subtypes,
and secondly, to determine similar biomarker sets for survival prediction in kidney and breast cancer patients.

Both case studies were addressed by using internal-external CV on TCGA \cite{hutter2018cancer} data. Additionally, the breast cancer case study includes an external validation with training
on TCGA data and testing on SCAN-B data \cite{Brueffer2018}.
The latter test is particularly important as it demonstrates that the proposed algorithm can also be utilized by training predictive models
for both classification and overestimation within a specific cohort (e.g. TCGA) and applying them in external cohorts (e.g. SCAN-B).

TCGA transcriptomic datasets were downloaded
with the curatedTCGAData R-package version 2.0.1 from assays of type RNASeq2GeneNorm \cite{ramos2020multiomic}.
The retrieved data consists of upper-quartile-normalized TPM values. They were log-transformed by applying $log_{2}(x+1)$.

The external gene expression-based transcriptomic dataset for breast cancer was obtained from the Gene Expression Omnibus (GEO) database (GSE96058)
collected from the SCAN-B consortium. It includes FPKM log-transformed gene expression profiles. This data was already log-transformed, and we
did not apply our own log-transformation to it. We applied for each value $x$ the function $max(x,0)$,
then we excluded the genes with less than 30\% non-zero values.

For our study, we utilized TCGA datasets specific to cancer types. For the breast cancer case study, we used the TCGA-BRCA dataset.
Additionally, for kidney cancer, we considered a compendium of TCGA datasets, which includes: KICH (Kidney Chromophobe),
KIRC (Kidney Renal Clear Cell Carcinoma), and KIRP (Kidney Renal Papillary Cell Carcinoma).

Each transcriptomic dataset was filtered by removing genes with zero variance and gene expression values were standardized before use.
For both TCGA and SCAN-B, cancer patient samples were categorized based on the PAM50 cancer subtype signature, 
which is used to determine specific molecular subtypes of breast cancer. 
The subtypes identified by PAM50 that were used for our experiments are: 
Luminal A (LumA), Luminal B (LumB), HER2-enriched (Her2), Basal-like (Basal, which is often referred to as triple-negative) and normal-like (Normal) cancer. 
For the task of classifying cancer subtypes in TCGA kidney cancer patients, we focused on distinguishing clear-cell renal cell carcinoma (ccRCC), 
chromophobe renal cell carcinoma (ChRCC), and papillary renal cell carcinoma, which was further divided into two subtypes based on recent studies identifying distinct clinical categories. 
These are referred to as PRCC T1 and PRCC T2. Additionally, we included samples of normal non-cancerous tissues. 
This classification system is based on a study published by Ricketts et al. \cite{ricketts2018}.
Overall survival data for TCGA kidney cancer is from Liu et al. \cite{liu2018}.

\mySupFig{tcga_ki_acc_pca}{First two principal components with subtypes of all the TCGA kidney samples.}
\mySupFig{tcga_ki_os_pca}{First two principal components with subtypes of the TCGA kidney samples with survival labels.}
\mySupFig{survival_stackplot}{TCGA kidney survival outcomes.}

The TCGA kidney dataset at the end of curation includes 5 classes and a total of 924 samples.
 \supFigRef{tcga_ki_acc_pca} shows the first two principal components (after standardization) and the
number of mRNA-based profiles for each kidney cancer subtype.
When the prediction of overall survival was included in the objectives, a smaller dataset of 793 samples was used in order to avoid the
unlabelled samples (\supFigRef{tcga_ki_os_pca}). This led to the removal of all the samples of normal (healthy) tissue,
because in TCGA they are collected from healty areas of cancer patients.
Additionally, two samples are excluded because they are missing the survival information.
\supFigRef{survival_stackplot} shows the distribution of the survival outcomes in time.

\mySupFig{tcga_br_pca}{First two principal components with subtypes of the TCGA breast samples.}
\mySupFig{scanb_pca}{First two principal components with subtypes of the SCAN-B samples.}

The TCGA breast dataset at the end of curation includes 5 classes and a total of 1081 samples.
\supFigRef{tcga_br_pca} shows the first two principal components (after standardization) and the
number of mRNA-based profiles for each breast cancer subtype in TCGA data.
The SCAN-B dataset at the end of curation includes 5 classes and a total of 2969 samples (\supFigRef{scanb_pca}).

\subsection{Experimental setup} \label{sec:setup}

Our tests include three datasets: TCGA kidney used for internal k-fold CV,
TCGA breast used for internal-external CV and for the training part of external validation,
and SCAN-B used for the testing part of external validation. The datasets and their preprocessing are
described in \supSecRef{datasets}.

Our MO problems include three objectives that measure the ability to classify the cancer subtypes,
the ability to predict the overall survival, and the feature set size. These objectives
are measured in the $[0, 1]$ interval as per assumption of the algorithms.
Subtype classification is measured with the balanced accuracy,
survival prediction with the c-index, while
the feature set size with the novel metric root-leanness.
It was measured with leanness in a previous study \cite{Cattelani2022}.
The leanness is a value between 0 and 1 with higher values for solutions with less features.
It is defined as $1/(n+1)$, with $n$ being the number of features.
The leanness decreases sharply as the feature set size increases;
consequently, the CHV is strongly impacted by the accuracy of the
smallest biomarkers that use 1 or 2 features.
Taking this into account, Cattelani et al. \cite{Cattelani2023}
provided an additional evaluation
of the same solution sets by using a different measure for the impact
of the number of features: $1/(\sqrt{n}+1$).
This soft-leanness decreases less sharply as the set size increases.
the root-leanness is even smoother than the soft-leanness and
is defined as the root of the leanness: $\sqrt{1/(n+1)}$.

The program performs a 3-fold CV repeated 3 times
for internal k-fold CV on
the kidney dataset, while, for uniformity with previous works
\cite{Cattelani2022, Cattelani2023}, a 5-fold CV on the breast dataset.
The MO optimizer used is the \algorithmName{} (Section 2.2),
wrapping NSGA3-CHS (\supSecRef{nsgaStar}) as both
tuning optimizer and main optimizer.
The hyperparameters of NSGA3-CHS used as main algorithm are population 500, generations 500, and
3 folds used for the evaluation of the individuals. The hyperparameters
related to the tuning phase of the \algorithmName{} are set as explained
in \supSecRef{overhead}. Each test is repeated 6 times, one for each of the
5 regressors described in \supSecRef{regressors}, and one for the zero regressor
(equivalent to the unadjusted NSGA3-CHS).

In this study, for the task of cancer subtype classification, NB and SVM were selected as the inner models. 
Meanwhile, Cox was employed as the inner model for the task of cancer survival prediction.
The SVM uses the Radial basis function kernel, balanced class weight,
and l2 regularization parameter $C=1$.

The considered experimental setups are listed in Table 1.

In each k-fold CV execution, samples are stratified based on the included objectives. 
For balanced accuracy, stratification is by cancer subtype, and for overall survival, 
by survival status and time (binned into high or low for evenly distributed death events). 
If both these objectives are present, stratification combines both criteria.

During each GA call, for classification objectives, 
features not passing an ANOVA test (p-value 0.05) are removed from the current training samples. 
For survival objectives, features that fail a Wald test (p-value 0.05) in a Cox regression are discarded. 
If both objectives apply, features failing both tests are dropped.

For each combination of validation type, datasets, objectives, adjusting regression model, 
and classification inner model, we calculated the MOPE and $P_{\Delta}$ to assess 
the difference between expected and actual performance on the test samples. 
The CHV was also computed as an overall performance indicator for the solution sets.
To the best of our knowledge it is
the only proposed generalization of both HV to CV scenarios
and of single-objective CV to MO \cite{Cattelani2023}. CHV takes
into account the differences between the performance expected
by the optimization algorithm and measured on
the test samples and preserves the HV appreciated properties,
in particoular the strict monotonicity \cite{Li2020}: if a set
of solutions is strictly better, the CHV is guaranteed to
be higher.

\section{Supplementary results}

\subsection{Meta-features correlation}\label{sec:correlation}

\mySupFig{fold_ki_acc_example}{
Example of scatter plot depicting solutions from CV on kidney cancer transcriptomics data.
MO optimization of balanced accuracy for subtypes prediction and root-leanness.
Horizontally, the number of features is shown for simplicity.
For each solution it is shown the performance measured in the inner CV, i.e.\@ the performance
expected by the optimizer, the performance of the model trained on the whole training set and tested
on the same set, and the performance of the same model on the testing set.
The number of features, the expected performance and the overestimation are correlated.}

\supFigRef{fold_ki_acc_example} shows an example of solution set performance after executing optimization
with NSGA3-CHS for kidney cancer subtypes biomarker discovery,
without any performance adjustment.
The optimizer explored the best trade-offs between balanced accuracy and feature set size, and used 3-fold CV internally to evaluate the solutions.
The overestimation (difference between the yellow and blue dots) of the balanced accuracy increases as the number of features increases.
Analogously, the overestimation increases when the estimation increases. Also the variance of the estimation,
computed by bootstrapping the samples, is correlated with the overestimation: the 9 p-values of the Kendall correlation for a
3-fold CV repeated 3 times are 0.008, 0.000, 0.068, 0.020, 0.006, 0.012, 0.003, 0.034, and 0.021.

\subsection{Effect of \algorithmName{} on solution sets in external validation}\label{sec:solution_sets}

In this section, we examine a case study that utilizes two distinct cohorts, TCGA breast and SCAN-B,
for external validation in biomarker discovery. 
The approach focuses on learning models and overestimation during biomarker identification within one cohort (like TCGA)
and then applying this knowledge to another cohort (such as SCAN-B). This shows the reliability of biomarker discovery across different cohorts.

As previously shown, in the external validation for classification of breast cancer subtypes the \algorithmName{}
with SVR as fitness adjustment regression model and SVM as inner classifier provides the best overall performance measured with the CHV,
while, keeping the SVM as inner classifier, using the RFReg it is possible to obtain the more accurate prediction of the performance on a new cohort,
with respect to both MOPE and $P_{\Delta}$.
The effect of the adjustment can be appreciated by looking at the solution sets presented in Fig.\@ 2.
The balanced accuracy expected by the optimizer ("internal cv'' in the figures) is monotonically increasing
with the number of features because the solutions returned by the algorithm are non-dominated.
The estimation error, i.e. the vertical distance between the balanced accuracy expected by the optimizer
using only the training samples and the balanced accuracy measured on the external dataset, is lower when using
the \algorithmName{}.

\section{Technical terms}\label{sec:terms}

The following is a list of the less common technical terms used in this paper.
When they can have more than one possible meaning, we specify the one to which we refer to.

\begin{description}

\item[Abstract class.]
A high-level structure for a group of subclasses that share common methods or attributes, but the implementation of these methods may vary across subclasses.

\item[ANOVA test.]
The ANOVA (Analysis of Variance) test is a statistical method used to compare the means of two or more groups of data sets to determine the extent of their difference.

\item[Applicability domain.]
The characteristics of the training set on which the model has been developed, and for which it is applicable to make predictions for new data.

\item[Balanced accuracy.]
A machine learning metric for binary and multi-class classification models, adjusted on imbalanced datasets by taking the average of sensitivity (true positive rate) and specificity (true negative rate).

\item[Binary tournament.]
A binary tournament in genetic algorithms is a selection method where two individuals are randomly chosen from the population, and the one with the better fitness score is selected for the next generation.

\item[Binned.]
A process that transforms continuous numerical variables into discrete categorical ``bins''.

\item[Biomarker.]
A measurable characteristic that reflects the physiological or pathological state of an organism, and these characteristics can be used to diagnose diseases, monitor treatment outcomes, or predict disease progression
\cite{ng2023benefits}.

\item[Biomarker discovery.]
The process of identifying and validating new biomarkers,
often using machine learning techniques to analyse high-dimensional biological data \cite{Diaz2022}.

\item[Biomarker model.]
A model that uses biomarker data to make predictions.

\item[Bootstrap.]
A statistical method used in machine learning for estimating the skill of machine learning models by resampling a dataset with replacement.

\item[Cancer subtypes.]
The categorization of specific cancers into distinct groups based on molecular or other characteristics.

\item[Concordance index.]
A measure of the predictive accuracy of a survival model, representing the probability that,
for a randomly selected pair of individuals, the one who experiences the event of interest first has a higher risk score.

\item[Cross hypervolume.]
Overall performance indicator for evaluating the solution set produced by a multi-objective optimizer using cross-validation.
It generalizes both the hypervolume to cross-validation scenarios and single-objective cross-validation to multi-objective problems \cite{Cattelani2023}. It preserves the hypervolume appreciated properties, in particular the strict monotonicity \cite{Li2020}: if a set of solutions is strictly better, the cross hypervolume is guaranteed to be higher.

\item[Cross-validation.]
A technique where a dataset is divided into two sections, one used to train a model and the other used to validate the model's performance.

\item[Decision maker.]
In the context of multi-objective problems, refers to the entity (which could be a person, a group of people, or an algorithm) that, once the set of potential solutions is identified, often represented by a Pareto front, selects the most suitable solution based on its preferences \cite{Roijers2017,Kaliszewski2018}.

\item[Estimate.]
A measure that quantifies how well a model is expected to perform on unseen data, often determined through techniques like cross-validation or using a separate test dataset.

\item[Estimation error.]
The difference between the estimated performance of a model (based on training or validation data) and the actual performance of the model when applied to new, unseen data.

\item[Evaluation.]
The process of assessing the performance of a model using certain metrics, which helps to understand how well the model generalizes to unseen data.

\item[Expected performance.]
The anticipated accuracy or effectiveness of a model in making predictions on new, unseen data.

\item[External set.]
A dataset that is different from the one used to train models, not just a different part of the same dataset.

\item[External validation.]
The process of testing a model's performance using a dataset that was not used in the training process
(not just a different part of the same dataset),
providing a more realistic assessment of how the model will perform on new, unseen data \cite{Ramspek2020}.

\item[Family of functions.]
Functions with a common definition that is parametric. The parameters can be other functions, like in the case of the CHV \cite{Cattelani2023}.

\item[Feature selection.]
The process of selecting a subset of relevant features (variables, predictors).

\item[Fitness estimation.]
The evaluation of a model's performance or the suitability of a solution in an optimization problem.

\item[FPKM.]
 FPKM (Fragments Per Kilobase Million) is a normalization method for quantifying gene expression from RNA-seq data, taking into account the effects of both sequencing depth and gene length.

\item[Generations.]
In genetic algorithms, a generation refers to one iteration of the algorithm, where a population of solutions is evolved towards better solutions.

\item[Genetic algorithm.]
An algorithm inspired by the process of natural selection, used to generate high-quality solutions to optimization and search problems by relying on biologically inspired operators such as mutation, crossover, and selection \cite{Alhijawi2023}.

\item[Hierarchical sort.]
A sorting process where data is organized in a hierarchical manner, meaning it's arranged according to levels of importance or significance.

\item[Hyperparameter.]
A parameter that is set before the learning process begins and can directly affect how well a model trains.
The set of features that are used is also a hyperparameter.

\item[Hyperparameter configuration.]
The specific set of hyperparameters used in a machine learning model.

\item[Hyperparameter tuning.]
The process of selecting the optimal values for a machine learning model’s hyperparameters.

\item[Hypervolume.]
In multi-objective optimization, the hypervolume indicator maps a solution set to the measure of the region dominated by that set \cite{guerreiro2021}.

\item[Individual.]
In the context of genetic algorithms, an individual refers to a candidate solution to an optimization problem,
which is analogous to a chromosome \cite{Alhijawi2023}.

\item[Inner model.]
A model to which a wrapper algorithm delegates the training and prediction for the considered hyperparameter configurations,
which includes the feature sets.

\item[K-fold cross-validation.]
A process where a dataset is randomly partitioned into $k$ equal sized subsets, and a model or optimizer is trained on $k-1$ subsets and validated on the left-out subset, with this process repeated $k$ times to ensure each subset is used for validation exactly once.

\item[Left-out samples/set.]
Data points that are left out of the training set and used as the test set in cross-validation.

\item[mRNA.]
Messenger ribonucleic acid.

\item[Meta-feature.]
A property or characteristic that can be used to guide the algorithm selection or hyperparameter tuning process.

\item[Model selection.]
The process of choosing one model from a set of candidate models based on their performance on a given dataset.

\item[Molecular biomarker.]
A specific molecule or set of molecules used as features in a model to help with the diagnosis or prognosis of diseases or disorders \cite{li2022data,Diaz2022}.

\item[Molecular features.]
The specific characteristics or properties of molecules that are used as input data for models.

\item[Molecular subtype.]
Subtype (of a disease) defined on the basis of molecular characteristics.

\item[Moving target.]
A problem where the objective or the data changes over time, requiring the model to adapt.

\item[Multi-objective feature selection.]
A process that aims to select the optimal subset of features considering multiple objectives, often involving trade-offs between conflicting objectives \cite{AlTashi2020}.

\item[Multi-objective problem.]
A type of problem that involves optimizing multiple conflicting objectives simultaneously, returning a set of solutions that approximates as much as possible the Pareto front.

\item[Mutation operator.]
In genetic algorithms, a mutation operator is used to introduce randomness and maintain diversity in the population by making small random changes in the individuals.

\item[Non-dominated.]
A solution such that no other solution is equal or better with respect to all objectives, and strictly better with respect to a least one objective.

\item[Non-dominated front.]
A set of solutions that are not dominated by any solution.

\item[Objective.]
A specific goal that the models are designed to achieve, such as maximizing classification accuracy or survival prediction.

\item[Objective function.]
A mathematical function used to quantify the performance of a solution with respect to an objective.

\item[On-line learning.]
A learning method where data becomes available in a sequential order and is used to update the best predictor for future data at each step.

\item[Optimizer.]
An algorithm that fine-tunes solutions with respect to the objectives of a problem.

\item[Overall survival.]
In the context of survival analysis in machine learning, refers to the time until the occurrence of death \cite{wang2019}.

\item[Overestimation.]
A performance estimation that is higher than the real one.

\item[Overfitting.]
Occurs when a machine learning model learns the detail and noise in the training data to the extent that it negatively impacts the performance of the model on new data.

\item[Partial derivative.]
A partial derivative of a function of several variables is its derivative with respect to one of those variables, with the others held constant.

\item[Pareto front.]
The set of solutions such that any other possible solution is dominated by them.

\item[Partitioned.]
Divided into subsets so that the subsets are disjunct and contain all the elements.

\item[Polymorphic.]
Polymorphism in programming refers to the ability of a variable, function, or object to take on multiple forms.

\item[Population.]
In the context of GAs, the collection of individuals that are currently considered by the algorithm.

\item[Primary sorting.]
In hierarchical sorting, the most important sorting criteria.

\item[Random search.]
A strategy that uses random combinations of hyperparameters to identify the optimal answer.

\item[Repeated CV.]
A technique where the CV process is repeated multiple times, providing a more robust estimate of the performance of a machine learning model.

\item[Secondary sorting.]
In hierarchical sorting, the second most important sorting criteria.

\item[Selected model.]
The model that gets selected from a collection of candidate models.

\item[Single-objective problem.]
A problem where the goal is to optimize a single objective.

\item[Solution.]
In the context of optimization algorithms, a possible solution to the optimization problem. It may be part of the final solution set, or a candidate solution during the execution of the algorithm.

\item[Stratified bootstrap.]
A resampling technique where the dataset is divided into strata and samples are drawn independently from each stratum.
The number of samples from each strata is relative to the strata prevalence.

\item[Survival analysis.]
A subfield of statistics used in machine learning to analyze and model time-to-event data, dealing with censoring \cite{wang2019}.

\item[Survival status.]
In survival analysis, whether an event of interest (such as death) has occurred \cite{wang2019}.

\item[Test performance.]
The evaluation on a test set of a machine learning model's ability to make accurate predictions when presented with unseen data.

\item[Tournament.]
In the context of GAs, a tournament involves running several ``tournaments'' among a few individuals (or ``chromosomes'') chosen at random from the population.

\item[TPM.]
TPM (Transcripts Per Kilobase Million) is a normalized measure of gene expression that first normalizes for gene length and then for sequencing depth,
making it easier to compare the proportion of reads that mapped to a gene in each sample.

\item[Train performance.]
The evaluation of a machine learning model's ability to learn and make accurate predictions on the training dataset.

\item[Training, validation, and test paradigm.]
A process where the dataset is divided into three subsets: training set (used to train the models), validation set (used to select the models), and test set (used to evaluate the final models' performance).

\item[Upper-quartile-normalized.]
A normalization method where the gene counts are divided by the upper quartile of counts different from zero in the computation of the normalization factors associated with their sample.

\item[Wald test.]
A hypothesis test done on the parameters calculated by the Maximum Likelihood Estimate (MLE) to check if the value of the true input parameters has the same likelihood as the parameters calculated by MLE.

\item[Wrapper algorithm.]
An optimization algorithm that considers the hyperparameter tuning as a search problem, where different combinations are prepared, evaluated, and compared to other combinations.

\end{description}

\bibliography{biblio}